# Band-Like Transport and Cation Off-Centring in Ag/Bi-Based Solar Absorbers

Yi-Teng Huang,[1,2,†] Yixin Wang,[1,†] Georgia Fields,[1] Peixi Cong,[3] Yongjie Wang,[1] Jack E. N. Swallow,[3,4] Avari Roy,[5] Jack M. Woolley,[6,7] Victoria Rotaru,[8,9] Maxim Guc,[8] Lars van Turnhout,[10] Mohamed Aouane,[11] Emmanuelle Suard,[12] Dominik Kubicki,[13] Alejandro Pérez-Rodríguez,[8,14] Aditya Sadhanala,[5] Akshay Rao,[10] Dennis Friedrich,[15] Robert S. Weatherup,[3] Simon J. Clarke,[1] Seán R. Kavanagh,[16,*] Robert L. Z. Hoye[1,*]

1. Inorganic Chemistry Laboratory, University of Oxford, South Parks Road, Oxford OX1 3QR, UK

2. Graduate Institute of Photonics and Optoelectronics and Department of Electrical Engineering, National Taiwan University, Taipei 10617, Taiwan

3. Department of Materials, University of Oxford, Parks Road, Oxford OX1 3PH, UK

4. Department of Chemistry, University of Manchester, Oxford Road, Manchester, M13 9PL, UK

5. Centre for Nano Science and Engineering, Indian Institute of Science, Bengaluru 560012, Karnataka, India

6. Department of Physics, University of Warwick, Coventry, CV4 7AL UK

7. Warwick Centre for Ultrafast Spectroscopy, University of Warwick, Coventry, UK

8. Catalonia Institute for Energy Research (IREC), 08930 Barcelona, Spain

9. Facultat de Fisica, Universitat de Barcelona (UB), 08028 Barcelona, Spain

10. Cavendish Laboratory, University of Cambridge, Cambridge, UK

11. ISIS Neutron and Muon Source, Science and Technology Facilities Council, Rutherford Appleton Laboratory, Didcot OX11 0QX, United Kingdom

12. Institut Laue-Langevin, 71 avenue des Martyrs - CS 20156 - 38042 GRENOBLE CEDEX 9, France

13. School of Chemistry, University of Birmingham, Edgbaston B15 2TT, United Kingdom

14. Departament d'Enginyeria Electrònica i Biomèdica, IN2UB, Universitat de Barcelona, 08028 Barcelona, Spain

15. Institute for Solar Fuels, Helmholtz-Zentrum Berlin für Materialien und Energie GmbH, Berlin, Germany

16. Yusuf Hamied Department of Chemistry, University of Cambridge, Cambridge, UK

† These authors contributed equally to this work

**Email:** sk2045@cam.ac.uk (S. R. K.), robert.hoye@chem.ox.ac.uk (R. L. Z. H.)

## Abstract

Ag(I)-Bi(III)-based semiconductors have gained substantial attention as nontoxic, stable alternatives to lead-halide perovskites for optoelectronics, but are widely limited by carrier localization, which severely restricts diffusion lengths. The most efficient Ag/Bi solar absorber is $AgBiS_2$, but diffusion lengths in nanocrystal films are <50 nm. Carrier localization in this rock-salt ($Fm\bar{3}m$) system is believed to arise from cation disorder, and so we herein investigate the layered cation-ordered analogue. Through beyond-DFT simulations combined with neutron and X-ray powder diffraction, we reveal that off-centring of $Ag^+$ and $Bi^{3+}$ cations is energetically-favoured in this cation-ordered phase. Despite local distortions in the $AgS_6$ and $BiS_6$ octahedra, band-like transport takes place, which, surprisingly, also occurs in the cation-disordered rock-salt phase when these materials are made as bulk powders. The cubic-phase powders have the same degree of cation disorder as the nanocrystals that have carrier localization, which suggests that extrinsic factors play a determining role. We ascribe the intrinsic band-like transport of both phases of $AgBiS_2$ to its close packing, ensuring high electronic dimensionality. These insights offer pathways for designing solar absorbers avoiding carrier localization limitations, and call for future efforts to enhance the efficiency of $AgBiS_2$ photovoltaics to focus on large-grained thin films, or improved nanocrystal surface passivation.

## Introduction

The past decade has witnessed a renaissance in solar materials discovery, driven by both the successes and limitations of lead-halide perovskites (LHPs). Many studies have aimed to emulate the exceptional optoelectronic properties of LHPs in alternative 'perovskite-inspired' materials (PIMs) that avoid issues with stability and toxicity[1, 2]. A common chemical motif found among PIMs is the A(I)-A'(III) combination of cations, whereby two Pb(II) cations are replaced with a monovalent and trivalent cation to maintain charge balance. Ag(I)-Bi(III) is a common pair because they are in stable oxidation states (unlike $Sn^{2+}$ or $Ge^{2+}$)[3], have low toxicity (unlike $Pb^{2+}$ or $Tl^+$), are not scarce (unlike In)[4] and have occupied semi-valence states ($s^2$ lone pair for Bi(III), $d^{10}$ sub-shell for Ag(I)) which can contribute to 'defect tolerance'[1, 5, 6]. Of the multitude of Ag(I)-Bi(III) solar absorbers investigated[7-13], the highest photovoltaic performance has been achieved with the cubic cation-disordered rock-salt phase of $AgBiS_2$, recently surpassing 10% power conversion efficiency (PCE)[14, 15]. Although rapid progress has been made over the past five years, the performance of $AgBiS_2$ solar cells, and other Ag(I)-Bi(III)-based PIMs, falls well below that of LHP photovoltaics[1], with a common critical limitation being carrier localization.

Carrier localization takes place due to small polaron or self-trapped exciton formation, as a result of strong carrier-phonon coupling. This phenomenon can severely limit photovoltaic performance by reducing charge-carrier mobilities, which then lowers diffusion lengths and the internal quantum efficiency (IQE) of photovoltaic devices[16, 17]. Carrier localization has been so widely found with Ag(I)-Bi(III) materials (*e.g.*, in $Cs_2AgBiBr_6$[18, 19] and Cu-Ag-Bi-I[20, 21] compounds) that it has been labelled a 'hallmark' of these PIMs[16, 22, 23]. However, the physical origins remain unclear, and it is unknown whether this limitation is intrinsic or can be overcome.

The highest PCEs for $AgBiS_2$ have been achieved with nanocrystals (NCs) of the cation-disordered rock-salt phase, where carrier localization has been shown to limit diffusion lengths to <50 nm (based on the mobility and recombination rate constants within each NC), such that the most efficient certified $AgBiS_2$ solar cells are comprised of nanocrystal films only around 30 nm thick[24, 25]. $AgBiS_2$ benefits from a high absorption coefficient >$10^5$ cm$^{-1}$ across the entire visible to near-infrared wavelength range[24], such that these ultrathin films are still adequately absorbing. However, the range of performant film thicknesses is narrow; making films thicker than 50 nm leads to reductions in PCE due to reduced short-circuit current densities and IQEs[26]. This limited thickness tolerance makes the future manufacturing of nanocrystal $AgBiS_2$ solar cells challenging.

Recently, it was found that the degree of carrier localization in $AgBiS_2$ nanocrystals could be reduced by controlling cation disorder[24, 27]. Cubic-phase $AgBiS_2$ has the rock-salt structure ($Fm\overline{3}m$ space group), comprised of cubic close packed $S^{2-}$ anions, with $Ag^+$ and $Bi^{3+}$ disordered on average over the octahedral holes because of their similar ionic radii ($Ag^+$: 129 pm, $Bi^{3+}$: 117 pm), as shown in Fig. 1a. $Ag^+$ and $Bi^{3+}$ ions tend to be inhomogeneously distributed in as-synthesized nanocrystals[24, 28], but can be homogenized (*i.e.*, the sizes of locally Ag- or Bi-rich regions become smaller) by post-synthetic heat treatment at up to 150 °C. Such cation homogeneity was found to reduce (but not eliminate) carrier localization[27], but heat treating these nanocrystals beyond 115 °C led to sub-gap states that also reduced performance down to <3% PCE [24]. Alternative routes to achieving band-like transport in these materials are highly desirable to improve the performance, reliability and manufacturability of $AgBiS_2$.

Beyond solar cells, controlling carrier localization is pivotal for improving the performance of thermoelectric materials, for which ternary chalcogenides, including $AgBiS_2$, have shown excellent potential[29-31]. High thermoelectric performance in this family of compounds is mostly attributed to their low lattice thermal conductivity and high Seebeck coefficients, most notably with recent reports of $AgSbTe_2$ [32]. Reducing carrier localization is essential for maximizing the power factor (through high electrical conductivity) and thus increase the thermoelectric figure-of-merit, $ZT$ [29].

Herein, we examined the layered cation-ordered phase of $AgBiS_2$. Like the rock-salt phase, the cation-ordered phase has a cubic close packed arrangement of $S^{2-}$ anions, but the Ag(I) and Bi(III) ions now occupy alternate layers of octahedral sites along the <111> direction (Fig. 1a, 1b). This layered cation-ordered analogue has rarely been studied, and there have been no detailed spectroscopic investigations into its charge-carrier kinetics, or computational analyses of carrier localization. Furthermore, there are disagreements over the structure adopted. Many previous reports claim $R\overline{3}m$ or $P\overline{3}m1$ space groups[28, 30, 33-36], with Ag(I) and Bi(III) both centred in their respective octahedra. However, a recent single-crystal determination from a naturally-occurring sample found the structure to be in space group $P3_221$ with off-centred Ag(I) and Bi(III)[37]. Recent computational investigations also suggest that Ag(I) is frustrated in an octahedral site, preferring a tetrahedral coordination instead ($P3m1$ space group proposed), but without experimental evidence[33]. Given these wide disagreements, it is critical to examine the structure of the cation-ordered phase of $AgBiS_2$, as the validity of computational characterisation and many experimental interpretations rely on correct identification of the atomic structure. In this work, we used a wide range of computational, diffraction and spectroscopic techniques to understand the local coordination geometry about Ag(I) and Bi(III), as well as the macroscopic structure of the cation-ordered phase. Local structural distortions can impact phonon transport, which is especially important for thermoelectric performance through its effect on lattice thermal conductivity. We therefore performed advanced quantum mechanical modelling of cation-ordered $AgBiS_2$ to unravel the underlying principles that control cation off-centring.

To understand carrier localization, we use optical pump terahertz probe (OPTP) spectroscopy,

surprisingly finding band-like transport in both cation-ordered and cation-disordered AgBiS$_2$ bulk powder samples, which are comprised of larger crystallites than the NCs. We also synthesized cubic-phase NCs (6 nm size) with a similar degree of cation disorder as the cubic powders, but found ultrafast carrier localization instead (1.18 ps$^{-1}$ localization rate), consistent with prior literature[27]. Through computational analyses, we find that band-like transport is expected for both phases of AgBiS$_2$, particularly when cation clustering is minimized in the cubic phase. This work reveals AgBiS$_2$ to intrinsically avoid the carrier localization limitations widely found in Ag(I)-Bi(III) PIMs, and that the observed self-trapping in AgBiS$_2$ NCs arises from extrinsic factors. These findings redirect future efforts with AgBiS$_2$ photovoltaics towards bulk materials, or nanocrystal passivation, and provide valuable insights for the design of Ag(I)-Bi(III) PIMs without carrier localization.

## Results

### *Synthesizing cation-ordered and disordered phases of AgBiS$_2$*

The cation-ordered phase of AgBiS$_2$ is more challenging to synthesize than the cation-disordered phase, and is therefore more rarely found in the literature. In the cubic phase, Ag(I) and Bi(III) occupy the same crystallographic site, with a probabilistic distribution, leading to higher configurational entropy[33]. Thus, whilst the layered cation-ordered phase has lower internal formation energy, the cation-disordered cubic phase has lower Gibbs free energy at high processing temperatures. To isolate the cation-ordered phase, we firstly obtained the cubic phase by melt synthesis at 850 °C (AgBiS$_2$ melts at 800 °C[38]), followed by 7 days of heat treatment at 140 °C (details in Methods). We verified through inspection of the powder X-ray diffraction (PXRD) patterns (Fig. 1c) that cubic-phase impurities were negligible. On initial inspection, the PXRD pattern appears to fit well to the Caswellsilverite phase ($R\bar{3}m$ space group), as reported by others[35, 36] (details in Supplementary Note 1, Supplementary Fig. 1-4).

These PXRD patterns were obtained on a laboratory X-ray source with conventional scan durations, as others have used previously. However, if we collect PXRD patterns with higher signal-to-noise ratio, together with neutron powder diffraction, we reveal symmetry lowering that is not observable with conventional scans, as we detail below.

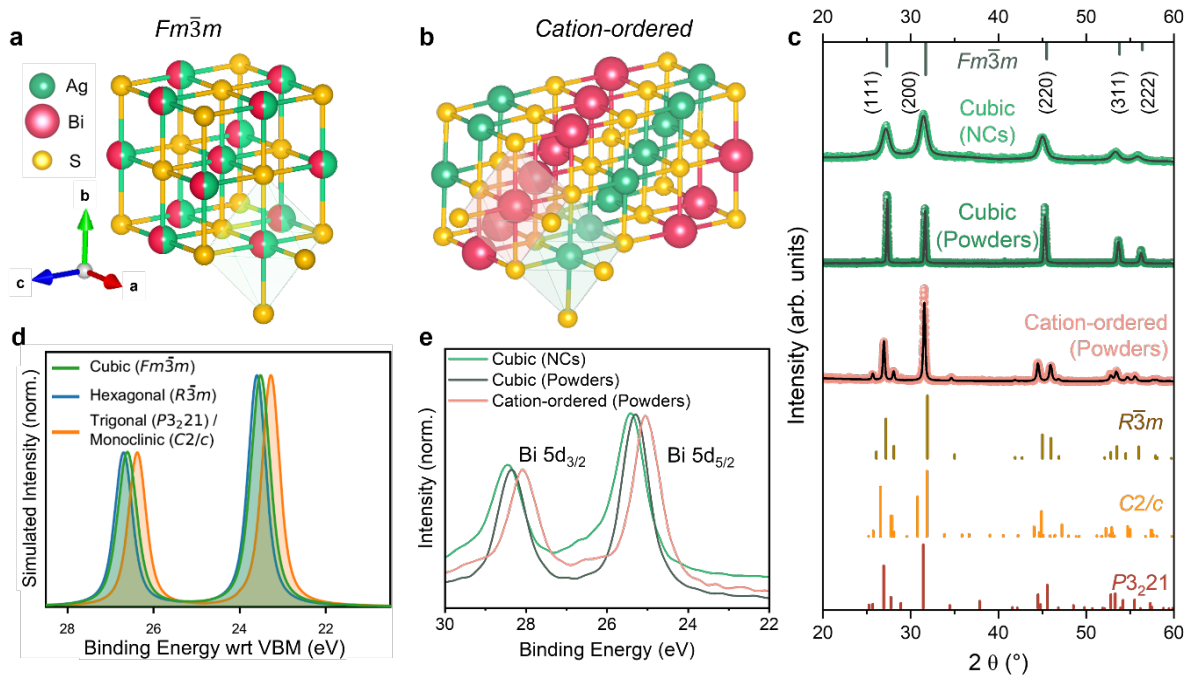

**Fig. 1 | Structures of cation-disordered and ordered phases of AgBiS₂.** Illustration of the **a** cubic rock-salt phase (space group: $Fm\bar{3}m$)[39] with homogeneous cation disorder, and **b** cation-ordered phase. This has a $R\bar{3}m$ space group if both cations are centred in their octahedra, or, with off-centring, form either a monoclinic $C2/c$ or trigonal $P3_221$ phase. Green, pink and yellow spheres refer to Ag, Bi, and S, respectively. **c** Comparative room-temperature X-ray diffraction pattern of cation-disordered cubic-phase nanocrystal, cubic-phase powder, and cation-ordered powder samples of AgBiS₂. Measurements are shown as spheres, while reference patterns of the $Fm\bar{3}m$ and three possible cation-ordered phases ($R\bar{3}m$, $C2/c$ and $P3_221$) are shown as sticks. Rietveld refinement profiles are shown in Supplementary Fig. 1. **d** Calculated X-ray photoemission spectra of Bi $5d$ semi-core levels in the three phases of AgBiS₂, with energies given relative to the valence band maximum (VBM) positions. **e** Measured X-ray photoemission spectra of cubic powders and nanocrystals, and cation-ordered powders of AgBiS₂. The spectra for the cubic NCs and cation-ordered powders show extra peaks centred at around 26.5 and 29 eV binding energy, which are attributed to surface oxide (*i.e.*, Bi-O[40]). Fits shown in Supplementary Fig. 8. Measured survey, valence band and C $1s$ spectra shown in Supplementary Figs. 9-10. Bi $5d$ spectra relative to VBM shown in Supplementary Figs. 11.

We also prepared cubic-phase powders by rapid quenching from 850 °C after melt synthesis, and compared these against cubic AgBiS₂ NCs prepared by hot injection and post-heat treated at 110 °C (Fig. 1c; Rietveld refinement in Supplementary Fig. 1). These NCs produced solar cells with a high PCE of 8.53% (Supplementary Fig. 5), showing that the NCs we produced for this work are state-of-the-art[24, 25]. We compared the homogeneity of cation disorder in these samples through the XRD peak positions (Table 1 and Supplementary Fig. 6, 7), as well as the binding energies of the Bi $5d$ core peaks in XPS measurements (Fig. 1d, e). The similarity in Bragg peak positions and XPS core levels between the cubic powder and NC samples suggests that they have similar degrees of cation disorder. This is because more homogeneous cation disorder leads to a reduced lattice parameter originating from shorter average cation–S bond lengths[24], and lower binding energies of Bi core levels caused by higher Madelung potentials[24]. The lattice parameters shown in Table 1 are close to previous reports for $AgBiS_2$ NCs post-heat treated at 115 °C (Table 1), which still exhibited carrier localization[27].

**Table 1.** Comparison of the X-ray diffraction peak positions, lattice parameters and XPS Bi $5d$ peak positions of cubic $AgBiS_2$ samples with different levels of cation disorder homogeneity. The lattice parameters of samples from this work were determined from Rietveld refinement. XPS core peak binding energies relative to the VBM obtained from Supplementary Fig. 11.

| Sample | XRD (200) Peak Position | Lattice Parameter (Å) | XPS Bi $5d_{3/2}$ Peak Position Relative to VBM (eV) | Ref |
|---|---|---|---|---|
| As-prepared (NC) | 31.14° | 5.737 | 28.50 | [24] |
| 110 °C (NC) | 31.49° | 5.709(1) | 28.36 | This work |
| 115 °C (NC) | 31.52° | 5.670 | 28.35 | [24] |
| 150 °C (NC) | 31.56° | 5.663 | 28.25 | [24] |
| Cubic powders | 31.60° | 5.667(4) | 28.16 | This work |

***Energetically-favoured cation off-centring in layered $AgBiS_2$***

To examine how changes in cation ordering affects the crystal lattice dynamics, we calculated the phonon dispersion curves of the cation-disordered $Fm\overline{3}m$ phase and idealised cation-ordered trigonal $R\overline{3}m$ phase of $AgBiS_2$ (Fig. 2a, b). Compared to the cation-ordered phase, the cubic phase has a broader distribution of phonon modes due to cation disorder, which we simulated here using a special quasi-random structure (SQS) supercell[41, 42]. Inelastic neutron scattering was unable to distinguish between the phonon dispersion curves of these two phases (Supplementary Fig. 12), but we were able to resolve differences from Raman spectroscopy. For all phases, the low-energy phonon modes below 2.5 THz are dominated by vibrations from the heavy Ag(I) and Bi(III) cations, whereas the high-energy modes peaking at approximately 7.5 THz originate from the lighter S anions (Fig. 2a, b). Although the low-energy phonon modes were below the accessible energy range of our Raman system, we could resolve the high-energy vibrational modes. The cubic-phase powders show a broader spectrum than the cation-ordered phase powders (Fig. 2c), consistent with the calculated phonon dispersion curves (Fig. 2a, b, d, e). The Raman peak at around 7.5 THz of NCs is broader than the bulk cubic powders (Fig. 2c), possibly due to greater energetic disorder further smearing out the S vibrational modes (discussed more later).

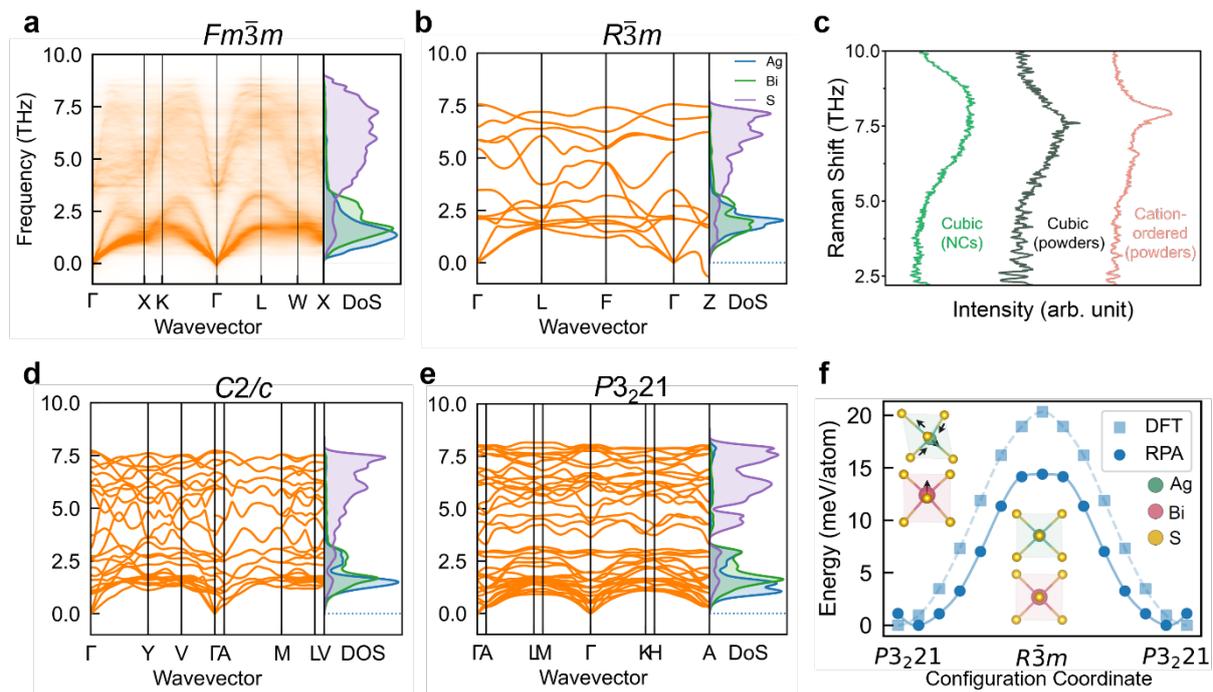

**Fig. 2 | Vibrational modes and local distortions in AgBiS₂.** Calculated phonon dispersion curves of **a** cubic ($Fm\bar{3}m$) and **b** hexagonal ($R\bar{3}m$) AgBiS₂. **c** Raman spectra of cation-disordered cubic NCs (light green), cubic powders (dark green) and cation-ordered powders (light pink). Calculated phonon dispersion curves of **d** monoclinic ($C2/c$) and **e** trigonal ($P3_221$) structures, with the atom-projected phonon density of states (DoS) alongside. **f** Configuration coordinate diagram of the minimum energy structural distortion path from $R\bar{3}m$ to $P3_221$ AgBiS₂, calculated using the Nudged Elastic Band (NEB) method with both the HSE06 hybrid DFT functional ("DFT") and (with fixed volumes) Random Phase Approximation (RPA). The slight offset in minima between HSE06 and RPA is due to slight differences in equilibrium geometries with these methods. The corresponding octahedral distortions are illustrated inset.

In examining the dynamic properties of the idealised $R\bar{3}m$ phase, we found that there is an imaginary phonon mode (*i.e.*, with frequency < 0) at the Z point in reciprocal space (Fig. 2b). This implies the presence of an energy-lowering, symmetry-breaking distortion for this phase, with an expansion of the unit cell. By applying the atomic displacements associated with this phonon mode and relaxing the structure without symmetry constraints, we obtained a monoclinic $C2/c$ phase which we calculated to be ~13 meV/atom lower in energy than the hexagonal $R\bar{3}m$ phase with the beyond-DFT Random Phase Approximation (RPA) method (Supplementary Fig. 13, Supplementary Table 1-9). This computationally-relaxed structure is similar to the recently-reported experimental $P3_221$ model for this phase[37], in that both Ag(I) and Bi(III) are off-centred (Fig. 3a; details in Supplementary Tables 1-4). Both phases have two short Ag–S bonds (~2.5 Å), along with two sets of longer Ag–S bonds (~3 Å). The main difference between the two phases is that $C2/c$ structure has the short Ag–S bonds *cis* to each other rather than *trans* ($P3_221$). Relaxing the geometry and computing the energy of the trigonal $P3_221$ phase we find it to be the lowest energy ordered structure for AgBiS$_2$, also being ~13 meV/atom lower in energy than the hexagonal $R\bar{3}m$ phase (and ~0.5 meV/atom lower energy than the $C2/c$ phase). Notably, we found that the energy differences between the AgBiS$_2$ polymorphs are surprisingly sensitive to the level of theory employed, as shown in Supplementary Tables 5, and so we employed computationally-intensive RPA calculations to obtain accurate predictions of the phase energetics. The minimum energy structural distortion pathway from $R\bar{3}m$ to $P3_221$ AgBiS$_2$ is mapped out in the configuration coordinate diagram shown in Fig. 2f (calculated with hybrid DFT and therefore showing a slightly larger energy difference of 21 meV/atom; Supplementary Table 5), illustrating how the AgS$_6$ and BiS$_6$ octahedra distort. There are no imaginary phonon modes in either of the $C2/c$ or $P3_221$ phases (Fig. 2 c, d), confirming no further *local* structural distortions to lower-energy states.

*Local cation coordination environment*

Having computationally predicted that local structural distortions occur in the cation-ordered phase (Fig. 3a), we experimentally analyzed the local environment about the cations in this phase in more detail.

Firstly, we performed extended X-ray absorption fine structure (EXAFS) measurements on powder samples, focussing on the local environment around Ag due to larger distortions around Ag than Bi. The magnitude of the Fourier-transformed $k^2$-weighted EXAFS measured at the Ag K-edge in transmission mode for cubic and cation-ordered $AgBiS_2$ are shown in Fig. 3b, where the corresponding fits are also displayed (see Methods for details). For the cation-ordered phase, we obtained a closer fit to the experimental data when our structural model included more than one Ag–S scattering path (Table 2). The multiple refined scattering lengths for cation-ordered $AgBiS_2$ powders are consistent with the multiple Ag–S bond lengths of the $C2/c$ or $P3_221$ structures that arise from structural distortions, and are qualitatively different from the single bond length in the $R\overline{3}m$ structure (Fig. 3c, Supplementary Fig. 14). EXAFS measurements therefore confirm the off-centring of Ag(I) in cation-ordered $AgBiS_2$, but cannot determine the exact polymorph. For the cation-disordered $AgBiS_2$, we found that the measurements were best fit with two Ag–S bond lengths (Table 2), matching our computationally-predicted radial distribution function (RDF; Supplementary Fig. 15). This is the result of the same preference for Ag(I) off-centring that we see in the ordered phase, though partially suppressed by the disordered Ag/Bi cation arrangement, which gives local distortions despite a macroscopically-cubic structure. Our computational simulations also explain the observation of short Ag-Ag distances in cation-disordered $AgBiS_2$, reported by Kesavan et al.[28], occurring in regions of higher Ag density where bonding locally becomes more $Ag_2S$-like.

**Table 2.** Bond length ($R$) and Debye–Waller factor ($\sigma^2$) for different Ag-S scattering paths extracted from the fittings to the Ag K-edge EXAFS spectra of cubic-phase (against $Fm\overline{3}m$ model) and cation-ordered (against $P3_221$ model) AgBiS$_2$ powders. Fitting of cation-ordered AgBiS$_2$ against $C2/c$ model is shown in Supplementary Fig.16 and Supplementary Table 10.

| Sample | Scattering Path | $R$ (Å) | $\sigma^2$ (Å$^2$) |
|---|---|---|---|
| Cubic | Ag-S$_1$ | 2.56(2) | 0.017(3) |
| | Ag-S$_2$ | 3.26(6) | 0.021(6) |
| Cation-ordered | Ag-S$_1$ | 2.53(3) | 0.013(3) |
| | Ag-S$_2$ | 2.85(8) | 0.025(9) |
| | Ag-S$_3$ | 3.17(12) | 0.035(12) |
| | Ag-Bi$_1$ | 3.69(11) | 0.052(22) |
| | Ag-Bi$_2$ | 3.76(11) | |
| | Ag-Bi$_3$ | 3.86(11) | |

For the local chemical environment about Bi(III), we used X-ray photoemission spectroscopy (XPS). We predicted from computations that the XPS spectrum for the $R\overline{3}m$ phase has a marginal increase in binding energy for the Bi $5d$ core level compared to the cubic phase (Fig. 1d) relative to their valence band maxima (VBMs). By contrast, we observed a decrease in binding energy (Fig. 1e), which is consistent with the simulated spectra for both the $C2/c$ and $P3_221$ phases. XPS therefore corroborates that the local geometry in the cation-ordered phase is not $R\overline{3}m$, but, like EXAFS, we are unable to conclude the exact cation-ordered phase from XPS results.

In view of the difficulty in determining whether the cation-ordered phase is $C2/c$ and $P3_221$, we used high-resolution PXRD, powder neutron diffraction (PND), and X-ray pair distribution function (PDF) analyses on our cation-ordered powder samples. These detailed analyses complement the single crystal report on a naturally-occurring specimen from which the $P3_221$ model was deduced[37]. PXRD patterns are dominated by the scattering from Bi, but offer good Bi/Ag contrast, while PND patterns have a greater relative contribution from the lighter S (scattering lengths: Bi = 8.53 fm; Ag = 5.92 fm and S = 2.85 fm) and the $Q$-independent scattering amplitude for neutrons means that the atomic positions may be determined with high precision. Upon close inspection, we found that the $R\overline{3}m$ model did not satisfactorily account for all reflections (Supplementary Figs. 17-20, Supplementary Table 11). On the other hand, the $P3_221$ model (constructed from the information in Ref. 37) accounts for all of the reflections, and comparison to computed patterns shows that no peaks are missing from the pattern, *i.e.*, there are no predicted reflections where the measured peak is indistinguishable from the background. A comparison of the $R\overline{3}m$ and $P3_221$ models can be found in Supplementary Fig. 21. The synchrotron X-ray data confirmed full Ag/Bi order, with 100(1)% ordering obtained from fitting these data. Analysis of PND data measured at 2 K showed the ordering to be similar (97(2)%; refer to Supplementary Table 12). Fitting the diffraction patterns of the cation-ordered phase with the $P3_221$ model produced significantly improved agreement factors (Fig. 3d) over refinements against the $R\overline{3}m$ or $C2/c$ models (Supplementary Table 11), with the refined lattice parameters and atomic coordinates in good agreement with those reported by Lissner *et al.* (Supplementary Tables 13-17)[37].

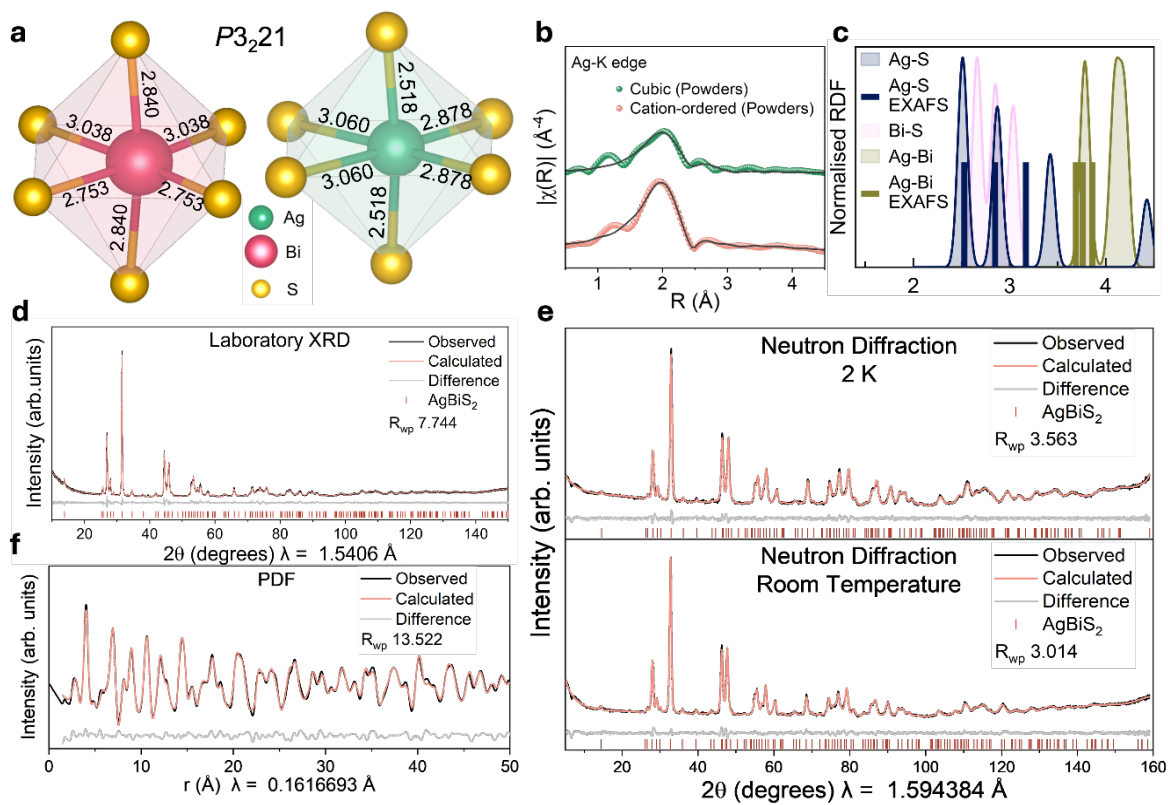

**Fig. 3 | Local structure in powder samples of AgBiS₂. a** Illustration of AgS₆ and BiS₆ octahedra in $P3_221$ AgBiS₂ model refined from powder neutron diffraction. All bond lengths in Å. **b** k²-weighted transmission Ag K-edge FT-EXAFS moduli (phase uncorrected) of the cation-disordered cubic and cation-ordered AgBiS₂ powders fit with the $P3_221$ model. Corresponding bond lengths given in Table 2. **c** Calculated radial distribution function (RDF) for cation-anion and cation-cation bonds in $P3_221$ AgBiS₂ structure. Gaussian smearing of width 0.05 Å was applied to mimic thermal broadening effects. Bond lengths extracted from experimental EXAFS are marked in bars. **d** High-resolution diffraction pattern of cation-ordered AgBiS₂ measured using a laboratory X-ray source with a long counting duration. Full comparative refinements of the three space groups ($R\bar{3}m$, $C2/c$ and $P3_221$) considered for cation-ordered AgBiS₂ using laboratory and synchrotron X-ray sources shown in Supplementary Figs. 17-18. **e** Neutron diffraction measured on D2B@ILL at low temperature (2 K) and room temperature of cation-ordered AgBiS₂. Comparative refinements of all space groups for cation-ordered AgBiS₂ are shown in Supplementary Fig. 19. **f** Rietveld plots of PDF data. All data from cation-ordered powders are fit to the $P3_221$ model. Comparative refinements of all space groups for cation-ordered AgBiS₂ are shown in Supplementary Fig. 20.

The refined $P3_221$ model from the PND measurements show distortions of the AgS₆ and BiS₆ units that are comparable with those obtained computationally (Fig. 3a, Supplementary Table 13-17), with the main difference being the length of the longest Ag-S bond pair (~3.4 Å from theory, ~3.1 Å from PND and single-crystal X-ray measurements[37], ~3.2 Å from EXAFS). In particular, the distortion around Ag⁺ produces two short (~2.5 Å) Ag-S bonds *trans* to one another, as is common for $d^{10}$ cations where these orbitals remain in the valence region. In addition, there are two intermediate (~2.9 Å) Ag-S bonds which are *cis* to one another, giving overall an environment tending towards tetrahedral, as shown in Fig. 3a. The distortion of the $Bi^{3+}$ octahedral environment is less significant (Table 3), mostly corresponding to an off-centring of Bi. The 2 K PND (Fig. 3e) model was similar to that obtained at ambient temperature, suggesting there is no structural change in this temperature range.

**Table 3.** Comparison of the distortions of $AgS_6$ (subscript Ag) and $BiS_6$ octahedra (subscript Bi) across the different polymorphs of $AgBiS_2$. The tabulated bond length distortion index ($D$) and bond angle variance ($\sigma(\theta)^2$) are obtained from calculated bond lengths and angles as shown in **Supplementary Table 1-4** using Supplementary Eq. S1-2.

|  | $D_{Ag}$ | $\sigma(\theta)^2_{Ag}$ | $D_{Bi}$ | $\sigma(\theta)^2_{Bi}$ |
|---|---|---|---|---|
| Cubic $Fm\overline{3}m$ | 0.08 | 22.0 | 0.06 | 10.4 |
| Trigonal $R\overline{3}m$ | 0 | 1.8 | 0 | 0.5 |
| Monoclinic $C2/c$ | 0.14 | 3.0 | 0.05 | 30.6 |
| Trigonal $P3_221$ | 0.10 | 122.2 | 0.05 | 22.0 |

As an additional probe of the local structure, X-ray PDF measurements were analyzed. The idealized $R\overline{3}m$ model was found to fit poorly across all length scales. The computed $C2/c$ model with local distortions around the cations performed much better at short distances, reflecting cation off-centring, but less well at longer distances out to 20.0 Å, and was therefore not able to adequately capture the long-range order (Supplementary Fig. 20). The $P3_221$ model produced the best agreement across all length scales (Fig. 3f, Supplementary Table 11), supporting the laboratory and synchrotron X-ray, the neutron diffraction data (Fig. 3d, e) and the single-crystal report[37].

### *Charge-carrier transport properties*

Having established the structural properties of the cation-disordered cubic and cation-ordered $AgBiS_2$ phases, we proceeded to compare their optoelectronic properties. Firstly, to establish the steady-state optical absorption spectra (Fig. 4a), we used photothermal deflection spectroscopy (PDS) to avoid excessive light scattering in reflectance measurements of thick powder samples, as detailed in Supplementary Note 3, with data shown in Supplementary Fig. 22. To account for energetic disorder in this system (*i.e.*, energetic fluctuations in band positions), we fit the absorption spectra with the bandgap fluctuation model (Supplementary Eq. S3-4), which models a distribution of bandgaps rather than a fixed bandgap with an Urbach tail below it[43]. The average bandgap obtained from this model cannot be directly compared to the calculated bandgaps in Table 4, owing to the inclusion of Gaussian broadening at the absorption onset. From this analysis, we found that the nanocrystals had the highest energetic disorder (0.415 eV), consistent with their high density of grain boundaries or surfaces. The cubic powders had the lowest energetic disorder (0.165 eV), consistent with its absorption onset being steepest (Fig. 4a), and these crystallite sizes being larger than the cation-ordered powders or NCs.

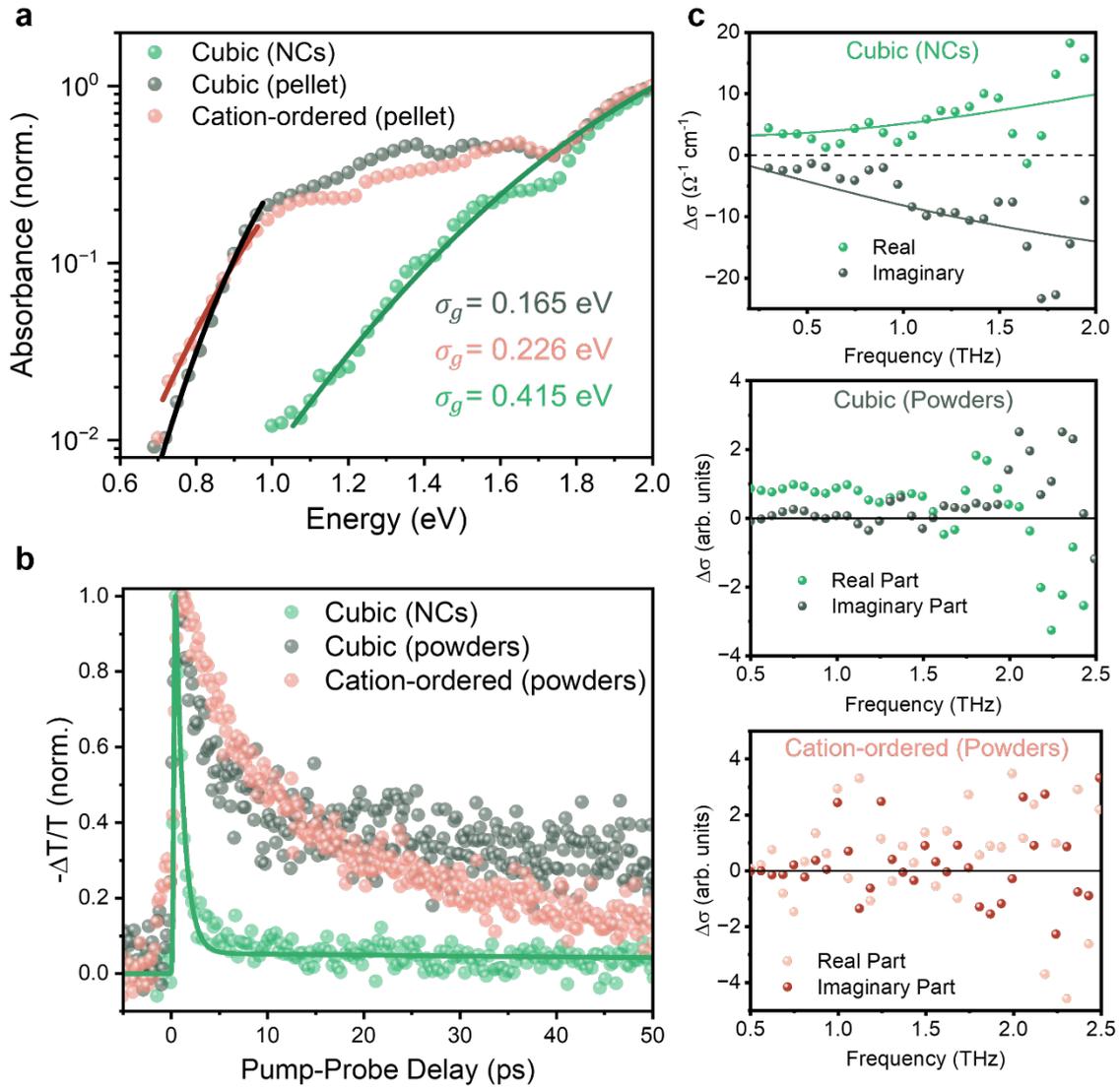

**Fig. 4 | Optoelectronic properties of AgBiS₂.** **a** Absorption spectra of cation-disordered cubic NCs, cubic pellets and cation-ordered AgBiS₂ pellets obtained from photothermal deflection spectroscopy. Solid lines represent the fits using the bandgap fluctuation model, with extracted energy disorder ($\sigma_g$) values of the three AgBiS₂ samples shown. The spectra were normalized to the absorbance at 2 eV. **b** Optical-pump-terahertz-probe (OPTP) kinetics of cation-disordered cubic NCs, cubic powders and cation-ordered powders AgBiS₂. **c** Complex photoconductivity spectra of the three AgBiS₂ samples acquired at 12 ps pump-probe delay time from OPTP. The modified Drude-Smith model fitting for cubic NCs is shown in solid lines.

**Table 4.** Calculated optoelectronic properties of AgBiS$_2$ in different space groups.

| | $Fm\bar{3}m$ | $R\bar{3}m$ | $C2/c$ | $P3_221$ |
|---|---|---|---|---|
| Bandgap (eV) | 0.83 | 0.86 | 1.26 | 0.86 |
| Direct Bandgap (eV) | 0.90 | 1.57 | 1.72 | 1.26 |
| $m^*_e$ | 0.24 | 0.18 | $0.20 - 0.31$ | $0.14 - 0.18$ |
| $m^*_h$ | 1.0 | $0.8 - 1.4^a$ | $0.3 - 0.7$ | $0.4 - 0.6$ |
| Relative Vibrational Free Energy (meV/atom) | 0.0 | 2.3 | 3.0 | 4.4 |
| $\varepsilon_\infty$ | 11.1 | 13.6 | 9.5 | 9.8 |
| $\varepsilon_0$ | 80.0 | 42.9 | 46.4 | 42.1 |
| $\omega_{LO}$ (THz) | 3.72 | 4.68 | 2.66 | 3.37 |
| $\alpha_e$ | 0.93 | 0.75 | 1.56 | 1.14 |
| $\alpha_h$ | 1.63 | 1.53 | 2.17 | 1.86 |
| $E_d^{CBM}$ (eV) | 1.24 | 2.96 | 2.63 | 2.28 |
| $E_d^{VBM}$ (eV) | 1.24 | 2.05 | 1.39 | 1.40 |
| $E_b$ (meV) | 0.3 | 0.9 | 0.7 | 0.7 |

$m^*_e$: harmonic mean effective mass of electrons (related to electronic conductivity); $m^*_h$: harmonic mean effective mass of holes (related to electronic conductivity); $\varepsilon_\infty$: high-frequency dielectric constant; $\varepsilon_0$: total static dielectric constant; $\omega_{LO}$: effective longitudinal optical (LO) phonon frequency; $\alpha_h$: Fröhlich coupling constant of holes; $\alpha_e$: Fröhlich coupling constant of electrons; $g_{ac}$: acoustic coupling constant; $E_d^{VBM}$: acoustic deformation potential of valence band maximum; $E_d^{CBM}$: acoustic deformation potential of conduction band maximum; $E_b$: Wannier-Mott binding energies in the hydrogenic model.

$^a$ The VBM in $R\bar{3}m$ AgBiS$_2$ shows heavy hole masses in all directions, but the next highest energy valence band at ~0.15 eV below the VBM, exhibits low hole effective masses of $0.24 - 0.34$ $m_0$.

To understand carrier localization in these materials, we used OPTP spectroscopy to probe the photoconductivity transients of all samples after excitation with 400 nm wavelength pump pulses (2.5 mJ cm$^{-2}$ fluence). The photoconductivity transient of cubic NCs decayed substantially within 5 ps, whereas those of the cubic and cation-ordered powders could persist up to 50 ps. The $t_{50}$ values, which refer to the pump-probe delay for the OPTP transient to decay by 50% of its original value, were determined to be 1.1, 9.9 and 10.6 ps for cubic NCs, cubic powders and cation-ordered powders, respectively. The $t_{50}$ values of the two powder samples exceed that of CuSbSe$_2$ (6.7 ps), which has been shown to exhibit band-like transport[44], whereas the $t_{50}$ of the cubic NCs is similar to Ag(I)-Bi(III) double perovskites that exhibit carrier localization[44]. We hence conclude that carrier localization, which manifests as an ultrafast decay in the photoconductivity transient, was present in the cubic NCs (consistent with a prior report[27]), but absent in both of the bulk powder samples. It should be noted that the differential transmission in OPTP depends not only on the photoconductivity, but also on the charge-to-photon branching ratio. The latter can be lowered through the formation of excitons, which can also lead to an ultrafast decay in the OPTP signal. We ruled out this possibility from the small calculated exciton binding energy ($E_b$; Table 4) and previous experimental reports[27, 45] (as detailed in Supplementary Note 3).

Prior works fit the photoconductivity transients in cubic $AgBiS_2$ NCs using a two-level model (Supplementary Note 3, Supplementary Eq. S5)[20]. This models a system in which photo-excited free charge-carriers initially form delocalized large polarons, before rapidly localizing into small polarons. To determine whether such a model is valid here, we first analyzed the complex photoconductivity spectra (Fig. 4c). The complex photoconductivity spectrum of cubic NCs exhibit a positive real component that increases with increasing frequency, along with a negative imaginary component that decreases with increasing frequency. This is consistent with the modified Drude-Smith model (Supplementary Eq. S6-7)[46], which is the phenomenological extension of the classical Drude model and commonly used to describe charge transport in confined materials. Fitting the complex photoconductivity with the modified Drude-Smith model, we obtained a $c_1$ coefficient of 1 and a high localization rate of 1.18 ps$^{-1}$ from the cubic NCs, both of which are features of strong carrier localization[46]. On the other hand, the complex photoconductivity spectra of the two powder samples – which do not show fast localization – deviate from the modified Drude-Smith model, and thus the two-level model cannot apply. Overall, the OPTP results show that fast carrier localization is only present in the cubic NCs, with no evidence of this occurring in the cubic or cation-ordered powders.

## Discussion

In this work, we set out with the hypothesis that the cation-ordered phase of $AgBiS_2$ would mitigate the carrier localization found in cubic NCs based on previous understanding of reduced mobilities and small polaron formation in cation-disordered $AgBiS_2$. But from our results, there was no evidence of carrier localization in powder samples of the cation-ordered or disordered phases, despite the latter having a similar cation distribution as the cubic NCs that did exhibit severe carrier localization. Moreover, we revealed that with cation ordering, it becomes energetically favourable for these cations to be located away from the centres of their octahedra, and for the distortions to be ordered on the long range, such that the structure adopted is different from the widely-assumed $R\overline{3}m$ or $P\overline{3}m1$ models. Given that the structural distortions from cation off-centring can affect phonon transport, with important implications for thermoelectrics, in this section we examine the underlying factors behind these structural distortions, and elaborate why band-like transport occurs. We especially focus on determining principles that could be generalizable to other materials systems that could enable higher-performance solar absorbers or thermoelectrics to be designed.

### *Delocalized charge-carriers*

To establish whether small polarons would intrinsically form in $AgBiS_2$, we performed first principles calculations, where we introduced an electron or hole into a supercell, distorted the structure, and relaxed this system to equilibrium. If small polarons form, the charge-carrier wavefunction would then be localized to within a unit cell, as previously found, for example, for $NaBiS_2$[47]. But as shown in Figs. 5a-c for electrons (and Supplementary Fig. 23 for holes), this was not the case. The $Fm\overline{3}m$ phase exhibited very minor localization due to local fluctuations in cation distributions, with electrons favouring Bi-rich regions and holes favouring S-rich regions. This implies that improving the homogeneity of cation disorder would indeed favour band-like transport, as indicated in previous work[27]. However, the absence of strong restriction of the electron/hole wavefunctions to within a unit cell suggests that the ultrafast carrier localization found in the cubic NC samples does not occur through intrinsic chemical or physical factors. By contrast, the idealised $R\overline{3}m$ and experimentally-observed $P3_221$ cation-ordered phases have fully delocalized wavefunctions. Thus, despite symmetry-breaking and distorted local environments in cation-ordered AgBiS$_2$, carrier localization is avoided because neither the $R\overline{3}m$ nor $P3_221$ phase intrinsically forms small polarons.

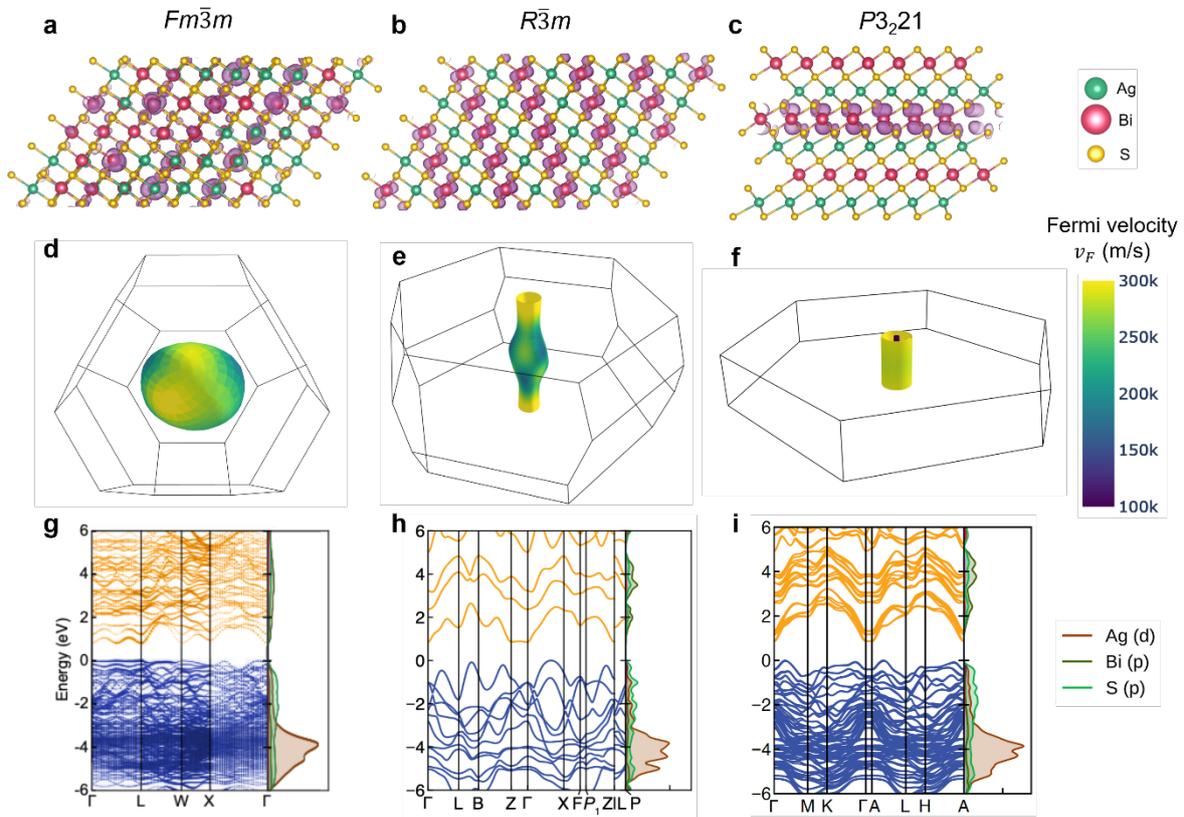

**Fig. 5 | Computational analysis of polarons, electronic dimensionality and electronic structure in AgBiS$_2$.** Charge densities (**a-c**) and Fermi surfaces (**d-f**) at the conduction band (0.1 eV above the CBM) for electrons in AgBiS$_2$ with the $Fm\overline{3}m$ (**a, d**), $R\overline{3}m$ (**b, e**) and $P3_221$ (**c, f**) phases, respectively. The Fermi surface colour in **d-f** corresponds to the Fermi velocity $v_F$. The electron wavefunctions (purple surfaces in **a-c**) are mostly concentrated across BiS$_6$ octahedra, as Bi $6p$ – S $3p$ antibonding states dominate the lower conduction band, while hole wavefunctions are mostly concentrated across S $p$ orbitals (Supplementary Note 4). Calculated electronic band structures and density of states (DOS) for the **g** $Fm\overline{3}m$, **h** $R\overline{3}m$ and **i** $P3_221$ phases, respectively.

Band-like transport occurs when there is an energy barrier between free and localized carriers, or when there is weak coupling between electrons and acoustic or longitudinal optical (LO) phonons. An energy barrier occurs when the electronic dimensionality is high (3D, or 2D with weak acoustic coupling)[16, 17]. By contrast, in a 2D system with strong acoustic coupling, or an electronically 1D or 0D system, small polarons would spontaneously form. Indeed, a common reason for the prevalence of carrier localization in Ag(I)-Bi(III) PIMs is that the electronic dimensionality is often 0D due to the energy mismatch in frontier orbitals between Ag(I) and Bi(III)[16]. However, we found the $Fm\overline{3}m$ phase of $AgBiS_2$ to be electronically 3D at the conduction and valence bands, as shown by the 3D charge density distribution and spherical Fermi surface in Fig. 5a, d and Supplementary Fig. 24a. For both the $R\overline{3}m$ and $P3_221$ structures of cation-ordered $AgBiS_2$, the conduction band is electronically 2D for the most part, as shown by the cylindrical Fermi surfaces (Fig. 5e, f) and charge density being discontinuous along only one dimension (Fig. 5b, c). The valence band is mixed 3D/2D for $R\overline{3}m$ and 2D for $P3_221$ (Supplementary Fig. 24b, c).

The high electronic dimensionality for $AgBiS_2$ can be ascribed to close packing, where all structures are based on, or derived from, a cubic closed packed arrangement of $S^{2-}$. In the $Fm\overline{3}m$, $R\overline{3}m$ and $P3_221$ phases, all metal-sulfide octahedra are edge-sharing, with strong continuous metal-S-metal interactions as a result. This close packing results in strong band dispersion, contributing to small band gaps, large high-frequency dielectric constants, high electronic dimensionality and low charge-carrier effective masses in these systems (Table 4; Fig. 5g-i). The high electronic dimensionality is in contrast to $Cs_2AgBiBr_6$, for example, where the $AgS_6$ and $BiS_6$ octahedra are only corner-sharing and thus spaced further apart, along with no continuous Ag-anion-Ag/Bi-anion-Bi interactions due to the checkerboard-like cation arrangement, such that the electronic dimensionality is instead 0D. In $Fm\overline{3}m$ $AgBiS_2$, the electronic dimensionality would only locally reduce in the presence of inhomogeneities, such as cation clustering to give Ag-rich or Bi-rich regions, which perturb the electronic potential and contribute to extended fluctuations in charge-carrier density[24, 47]. In layered cation-ordered AgBiS$_2$, strong intra-layer metal-anion-metal interactions leads to 2D electronic dimensionality (Fig. 5f and Supplementary Fig. 24c).

Further enhancing band dispersion in AgBiS$_2$ is the presence of anti-bonding interactions between the occupied Ag 4$d$ orbitals with S 3$p$ at the VBM that aids in lowering the hole effective masses[47]. We find that the Fröhlich coupling constants are in the weak-to-intermediate range (Table 4), which means that carrier localization cannot take place through electron coupling to LO phonon modes. The Fröhlich coupling constant ($\alpha$) is given by Eq. 1.

$$\alpha = \frac{e^2}{4\pi\epsilon_0}\left(\frac{1}{\epsilon_\infty} - \frac{1}{\epsilon_{\text{stat}}}\right)\sqrt{\frac{m}{2\omega_{\text{LO}}\hbar^3}} \qquad (1)$$

In Eq. 1, $\epsilon_0$ is the vacuum permittivity while $\epsilon_\infty$ and $\epsilon_{\text{stat}}$ are the calculated optical and static dielectric constants, respectively. $\omega_{\text{LO}}$ is the effective longitudinal optical (LO) phonon frequency, and $\hbar$ is the reduced Planck's constant. The low effective masses ($m$) therefore contribute to the low Fröhlich coupling constants. The dominant factor, however, is the relatively large high frequency dielectric constants $\epsilon_\infty$ (Table 4), which arise due to the small band gaps and high electronic densities of states at the band edges (leading to their ultra-strong absorption)[24, 47]. Both of these factors can again be linked to the dense structures and close-packing of these compounds, giving rise to high structural and electronic dimensionality, significant band dispersion and thus low band gaps and carrier effective masses. We note that these properties which favour delocalization are additionally related to the 'disorder tolerance' of this material, where prohibitive carrier trapping is avoided in the disordered cubic phase, akin to so-called 'defect tolerance' in lead-halide perovskites[48].

Moreover, as shown in Table 4, we find that the deformation potentials of these materials are all low, <3 eV in both the conduction and valence bands. Low deformation potentials correspond to polaronic localization being unfavourable, due to reduced electronic energy gain associated with lattice distortion. In comparison, $Cs_2AgBiBr_6$ has acoustic deformation potentials exceeding 13 eV, which is thought to contribute to carrier localization in this material[19], while our previous work on $CuSbSe_2$ found band-like transport and low deformation potentials (<7 eV)[44].

From this analysis, it is clear that the cubic cation-disordered phase of $AgBiS_2$ should only exhibit very minor carrier clustering, if at all, which accounts for the band-like transport of the cubic powders. The fact that cubic NCs exhibit carrier localization is likely due to extrinsic factors, particularly given the substantially higher grain boundary density in these materials compared to the bulk powder samples. We note that OPTP spectroscopy probes charge-carriers that move over a distance of 0.4 nm (Supplementary Eq. S8), well below the NC size of 6 nm[24, 27], and so the OPTP signal would be dominated by the more mobile carriers in the bulk of each NC. However, defects at surfaces may aggravate carrier localization. This implies that thin films with a larger grain size than the NCs should avoid small polaron formation and have longer charge-transport lengths. We therefore grew cubic-phase $AgBiS_2$ thin films by vapour-phase deposition, and found ultrafast localization to be absent (Supplementary Note 5, Supplementary Figs. 26-27). Indeed, $AgBiS_2$ films with a thickness of 265 nm have recently been reported to achieve >10% PCE[15], implying that these larger-grained films had diffusion lengths substantially exceeding the ~50 nm value for NCs[27], although carrier localization in these films were not probed until now. Our detailed spectroscopic and computational analyses explain why transport lengths in these bulk thin films could be longer than in NCs, showing that grain boundary engineering of thin films is a promising route for achieving efficient $AgBiS_2$ solar cells.

### *Rationalizing local structural distortions*

Turning then to the cation off-centring in trigonal $AgBiS_2$, we highlight that a key prediction from this finding is the splitting of Ag–S bond lengths into short (~2.5 Å) and long (~3 Å) pairs. We found that the distortion from $R\bar{3}m$ to $C2/c$ or $P3_221$ is similar to a second-order Jahn-Teller distortion, with the lower symmetry from $AgS_6$ octahedral distortion and Bi off-centring reducing the anti-bonding interactions between Ag $d$ / Bi $s$ and S $p$ in the upper valence band and giving a lower electrostatic energy. Using the Integrated Crystal Orbital Hamilton Population (ICOHP) shown in Supplementary Fig. 25, which provides a measure of covalent bonding, we found that the Ag-S and Bi-S ICOHP values become more negative (*i.e.*, stronger bonding), by 0.01 and 0.06 eV/bond respectively, in going from idealized, undistorted $R\bar{3}m$ to $P3_221$ $AgBiS_2$. Alongside, we found that this distortion lowers the electrostatic energy of the system due to the shorter cation-anion bonds and longer cation-cation distances as shown by the RDFs in Supplementary Fig. 14. Using the formal oxidation states (which overestimates the electrostatic energy change by neglecting screening), the calculated electrostatic energy decreased by 0.07 eV/atom with the distortion to $P3_221$.

Prior works attempting to rationalize bond anharmonicity in $AgBiS_2$ assumed this was due to the stereochemical expression of the Bi $6s^2$ lone pair[31]. While the Bi(III) lone pair certainly plays a role, we found greater distortion for Ag sites than Bi sites, as quantified by the larger $D$ and $\sigma^2$ values for $AgS_6$ in Table 3[28]. Furthermore, a recent theoretical work suggested that the ground state structure of $AgBiS_2$ would be $P3m1$, with Bi(III) in octahedral sites, but Ag(I) in tetrahedral sites due to greater covalency in the cation-anion bonds and the presence of a filled $d^{10}$ sub-shell[33]. However, this work noted the absence of experimental evidence[33], and such a structure is not supported by our data or that of others. In our work, we instead found from both experiment and calculations that Ag(I) occupies a distorted octahedron in $P3_221$, and that the $P3m1$ mixed-coordination phase is 23.1 meV/atom and 9.9 meV/atom higher in energy than the $P3_221$ and $R\bar{3}m$ phases, respectively (Supplementary Table 5). Notably, we find significant disagreement between DFT functionals on the relative energies of the $AgBiS_2$ phases, explaining this discrepancy[33] and necessitating the use of beyond-DFT methods (RPA) to obtain robust predictions of the ground-state structure in this work. Our results are consistent with observations that Ag(I) is unstable in an octahedral environment. In the $P3_221$ phase, the distorted $AgS_6$ octahedron has Ag(I) almost in a linear geometry with two very short bonds, which is often found when a filled $(n-1)d^{10}$ sub-shell is in close energetic proximity to the anion valence band (see Fig. 5) enabling anion-mediated mixing with the metal $ns$ valence states. In this case, there is a further distortion (Fig. 3a) to a site of point symmetry 2 ($C_2$).

Our findings of symmetry breaking, local structural distortions and the existence of several low-energy phases hint at the origin of ultralow lattice thermal conductivity in these compounds. Arising from frustrated environments and anharmonic covalent bonding, the resulting increased phonon scattering and slow heat transport has facilitated their adoption as high performance thermoelectric materials[31]. Notably, we expect these local structural distortions to occur in the wider $AgPnX_2$ family, where Pn is a pnictogen (Sb or Bi) and X is a chalcogen (S, Se and Te) such as $AgSbTe_2$, which exhibits low lattice thermal conductivity[49-51], and potentially in related $CuPnX_2$ and $AuPnX_2$ $d^{10}$ systems. This suggests an additional design rule to achieve low thermal conductivity for high thermoelectric performance is the presence of 'frustrated' atomic coordinations, where the tendency to distort to more favourable local configurations results in a loss of local structural symmetry and enhanced phonon scattering as seen in these compounds[32, 51]. Future work examining the impact of such effects – akin to octahedral distortions in lead-halide perovskites – on electronic structure in these compounds would help to further elucidate the impact of atom dynamics on optoelectronic performance in general.

## Conclusions

In conclusion, we found AgBiS$_2$ to intrinsically exhibit band-like transport in both the disordered rock-salt and cation-ordered polymorphs, with minor clustering of carrier wavefunctions in the cubic phase when the Ag/Bi distribution is inhomogeneous. We attribute this to the close-packed nature of these structures, where the close proximity of neighbouring edge-sharing AgS$_6$ and BiS$_6$ octahedra result in high electronic dimensionality, highly disperse bands, low deformation potentials, and strong electronic dielectric screening, along with high densities of states at band edges. These all disfavour ultrafast carrier localization by lowering acoustic and Fröhlich coupling constants, as well as making localized states less energetically favoured. The closely packed structure of AgBiS$_2$ contrasts to many other Ag(I)-Bi(III) PIMs explored, suggesting this to be a structural motif to target for designing efficient solar absorbers in this materials space. Furthermore, our OPTP measurements show a marked difference in carrier localization between bulk powders and NCs, despite both having the cubic-phase and similar levels of cation disorder, pointing to the crucial role of extrinsic factors, such as point defects, surfaces or grain boundaries. This redirects future efforts with AgBiS$_2$ photovoltaics toward bulk thin films with point/structural defect density, or careful passivation of surface defects in NCs, to improve efficiencies.

Furthermore, in examining the underexplored cation-ordered phase, we find off-centring of the Ag$^+$ and Bi$^{3+}$ cations from their octahedral sites, with significant consequences for structural and electronic properties (Table 4). The structure is in space group $P3_221$ rather than the widely-assumed $R\bar{3}m$ or $P\bar{3}m1$, which we could reveal only through high-level quantum-mechanical modelling along with detailed X-ray and neutron powder diffraction measurements with high-signal-to-noise ratios to reveal additional weak reflections, and synchrotron pair distribution function measurements. This supports a recent single-crystal structure report[37]. We propose that this cation off-centring is driven by the 'frustrated' nature of Ag(I) in an octahedral site, and off-centring Ag(I) and Bi(III) increases the overall bonding in the upper valence band. Such local structural distortions may be more generally found in materials with atomic species in frustrated coordination environments, enabling slow phonon transport, which is especially important for thermoelectric materials.

## Acknowledgements

We thank Dr. Johan Klarbring for helpful discussions regarding phase transformation timescales, Dr. Alex Squires for discussions regarding chemical origins of symmetry-breaking, Prof. A. L. Goodwin (University of Oxford) for providing access to INS beamtime, Prof. Dr T. Schleid (University of Stuttgart) for useful information regarding the origin of the crystal used for the cation-ordered phase[37], and Dr. Mark Isaacs for collecting the XPS data at the EPSRC National Facility for Photoelectron Spectroscopy ("HarwellXPS", EP/Y023587/1, EP/Y023609/1, EP/Y023536/1, EP/Y023552/1 and EP/Y023544/1). We thank the ILL for access to neutron beam time (doi:10.5291/ILL-DATA.EASY-1528) and the Diamond Light Source for access to I11 (allocation CY39378), I15 (allocation CY38927), and B18 (as part of the Energy Materials Block Allocation Group SP14239). Y.-T.H. and R.L.Z.H thank the Engineering and Physical Sciences Research Council (EPSRC) for funding (no. EP/V014498/2). Yi.W. and R.L.Z.H. thank the UK Research and Innovation for funding through a Frontier Grant (no. EP/X022900/1), awarded via the European Research Council 2021 Starting Grant scheme. R.L.Z.H. thanks the Henry Royce Institute for support through the Industrial Collaboration Programme, funded through EPSRC (no. EP/X527257/1). Yi. W. thanks funding from The Hilary and Galen Weston Foundation at Merton College Oxford, and from the Department of Chemistry at the University of Oxford. G.F would like to acknowledge funding from the EPSRC Centre for Doctoral Training in Inorganic Materials for Advances Manufacturing (IMAT), no. EP/Y035569/1. This research received funding from the European Research Council (ERC) under the European Union's Horizon 2020 research and innovation programme (EXISTAR, grant agreement No. 950598) and the UKRI through a Future Leaders Fellowship (MR/V024558/1). Av.R. and A.S. express gratitude to DST, SERB, MoE (MHRD), MeitY, QURP (Government of Karnataka), DST-SERD (DST/TMD/CERI/RES/2020/49), MoE-STARS (STARS-2/2023-0736), MoE SPARC-UKIERI (SPARC-UKIERI/2024-2025/P3975) and the Pratiksha Trust for their generous funding and support for this research. Av.R. and A.S. also extend our thanks to the Micro & Nano Characterization Facility (MNCF) at CeNSE for providing access to advanced characterization facilities facilitating the PDS measurements. The authors would like to acknowledge the University of Warwick Research Technology Platform, Warwick Centre for Ultrafast Spectroscopy, for use of Optical Pump Terahertz probe spectrometer in the research described in this paper. Experiments at the ISIS Neutron and Muon Source were supported by beamtime allocation RB2310425 from the Science and Technology Facilities Council. Data is available here: https://doi.org/10.5286/ISIS.E.RB2310425. Authors from IREC belong to the MNT-Solar Consolidated Research Group of the "Generalitat de Catalunya" (ref. 2021 SGR 01286) and are grateful to European Regional Development Funds (ERDF, FEDER Programa Competitivitat de Catalunya 2007-2013). V.R. acknowledges the support of the predoctoral program AGAUR-FI ajuts (2023 FI-1 00436) Joan Oro´ of the Secretariat of Universities and Research of the Department of Research and Universities of the Generalitat of Catalonia and the European Social Plus Fund. M.G. acknowledges the financial support from MCIN/AEI/10.13039/501100011033 and from FSE+ within the Ramón y Cajal (RYC2022-035588-I) program. S.J.C. thanks EPSRC for support through grant EP/T027991/1. R.L.Z.H.

also thanks the Royal Academy of Engineering for funding via the Senior Research Fellowships scheme (no. RCSRF2324-18-68).

## Author Contributions

R. L. Z. H. and Y.-T. H. conceived of this project. Y.-T. H. and Yi.W. synthesised AgBiS$_2$ powders and bulk thin films, performed lab-based XRD, and analysed PDS and OPTP, supervised by R.L.Z.H. G.F. synthesised the sample for neutron powder diffraction and analysed the diffraction and PDF data with S.J.C. P.C. and J.E.N.S. measured and analysed EXAFS results, supervised by R.S.W. Yo.W. synthesised AgBiS$_2$ NCs and fabricated photovoltaic devices. Av.R. conducted and analysed PDS, supervised by A.S. J.M.W. performed OPTP. V.R. and M.G. measured and analysed Raman data, supervised by A.P.-R. L.V.T. assisted PDS measurements, supervised by Ak.R. M.A. performed INS. E.S. performed PND. D.F. assisted OPTP measurements. S.R.K. performed the calculations and analysis. All authors wrote and edited the paper together.

## Competing Interests

The authors declare no competing interests

Supplementary Information for:

# Band-Like Transport and Cation Off-Centring in Ag/Bi-Based Solar Absorbers

Yi-Teng Huang,[1,2,†] Yixin Wang,[1,†] Georgia Fields,[1] Peixi Cong,[3] Yongjie Wang,[1] Jack E. N. Swallow,[3,4] Avari Roy,[5] Jack M. Woolley,[6,7] Victoria Rotaru,[8,9] Maxim Guc,[8] Lars van Turnhout,[10] Mohamed Aouane,[11] Emmanuelle Suard,[12] Dominik Kubicki,[13] Alejandro Pérez-Rodríguez,[8,14] Aditya Sadhanala,[5] Akshay Rao,[10] Dennis Friedrich,[15] Robert S. Weatherup,[3] Simon J. Clarke,[1] Seán R. Kavanagh,[16,*] Robert L. Z. Hoye[1,*]

[1.] Inorganic Chemistry Laboratory, University of Oxford, South Parks Road, Oxford OX1 3QR, UK

[2.] Graduate Institute of Photonics and Optoelectronics and Department of Electrical Engineering, National Taiwan University, Taipei 10617, Taiwan

[3.] Department of Materials, University of Oxford, Parks Road, Oxford OX1 3PH, UK

[4.] Department of Chemistry, University of Manchester, Oxford Road, Manchester, M13 9PL, UK

[5.] Centre for Nano Science and Engineering, Indian Institute of Science, Bengaluru 560012, Karnataka, India

[6.] Department of Physics, University of Warwick, Coventry, CV4 7AL UK

[7.] Warwick Centre for Ultrafast Spectroscopy, University of Warwick, Coventry, UK

[8.] Catalonia Institute for Energy Research (IREC), 08930 Barcelona, Spain

[9.] Facultat de Fisica, Universitat de Barcelona (UB), 08028 Barcelona, Spain

[10.] Cavendish Laboratory, University of Cambridge, Cambridge, UK

[11.] ISIS Neutron and Muon Source, Science and Technology Facilities Council, Rutherford Appleton Laboratory, Didcot OX11 0QX, United Kingdom

12. Institut Laue-Langevin, 71 avenue des Martyrs - CS 20156 - 38042 GRENOBLE CEDEX 9, France

13. School of Chemistry, University of Birmingham, Edgbaston B15 2TT, United Kingdom

14. Departament d'Enginyeria Electrònica i Biomèdica, IN2UB, Universitat de Barcelona, 08028 Barcelona, Spain

15. Institute for Solar Fuels, Helmholtz-Zentrum Berlin für Materialien und Energie GmbH, Berlin, Germany

16. Yusuf Hamied Department of Chemistry, University of Cambridge, Cambridge, UK

† These authors contributed equally to this work

*Email: : sk2045@cam.ac.uk (S. R. K.), robert.hoye@chem.ox.ac.uk (R. L. Z. H.)

# Table of Contents

**Supplementary Note 1 | Structural analysis of cation-disordered and ordered phases of AgBiS₂**

To quantify the phase-purity of the samples prepared, we used Rietveld refinement to fit the measured XRD patterns (Fig. 1c). We used the Schapbachite phase ($Fm\overline{3}m$ space group; ICSD coll. code: #42583) as the reference pattern to fit the cubic NC and powder samples, and all peaks were accounted for, with a weighted profile R-factor ($R_{wp}$) of 8.11 for the NCs (Supplementary Fig. 1a), and 11.53 for the powder samples (Supplementary Fig. 1b). The diffraction peaks of NCs also have lower intensities and broader widths than those of the powder sample, consistent with the smaller size of the NCs (Supplementary Fig. 1a).

Unlike the cubic phase, Ag(I) and Bi(III) occupy different octahedral sites in the cation-ordered phase, forming alternating layers of Ag(I) and Bi(III) layers along the <111> direction (Fig. 1b). Two possible space groups were initially thought to exist: $P\overline{3}m1$ (Matildite)[1] or $R\overline{3}m$ (Caswellsilverite), both of which have a cubic close-packed S²⁻ sub-lattice. In going from the $R\overline{3}m$ into $P\overline{3}m1$ phase, the one S position is split into three, and one position for each metal atom is also split into two. This transformation was already seen in the selenide analogue at low temperatures, as reported by Geller et al.[1] who reported that at low temperatures, the $R\overline{3}m$ phase transformed into the $P\overline{3}m1$ phase for the selenide analogue. Both structures are shown in Supplementary Fig. 2. The key difference in the reference patterns of these two phases is that $P\overline{3}m1$ has a (001) peak at 4.63° 2$q$ when using Cu K$_a$ radiation ($l$ = 1.54 Å)[1], whereas the $R\bar{3}m$ phase does not. As shown in Supplementary Fig. 3, we did not see any peak at 4.63° in our cation-ordered powder samples. Instead, the XRD pattern of our cation-ordered powder samples can be well refined by the Caswellsilverite structure ($R_{wp}$ of 11.85, Supplementary Fig. 1c), showing that the low-resolution XRD pattern can be fit to the $R\bar{3}m$ phase on average. In addition, neutron data collected at 2K (Fig. 3e) does not fit well to the $P\bar{3}m1$ phase, although the low angle peak described above cannot be seen due to limitations of the instrument. Synchrotron XRD patterns collected at 100 K (as part of the experiment shown in Supplementary Fig. 4), do not show any sign of a change to a low temperature $P\bar{3}m1$ phase either.

The transformation from the cation-ordered phase into the disordered phase is also possible at higher temperatures, as the lattice parameters would be expanded, and the silver/bismuth cations might move off their original sites to increase entropy. As displayed in Supplementary Fig. 4, it can be indeed seen that such a phase transformation occurred at 496 K, in close agreement with the temperature (483 K) reported by Ramdohr et al[1].

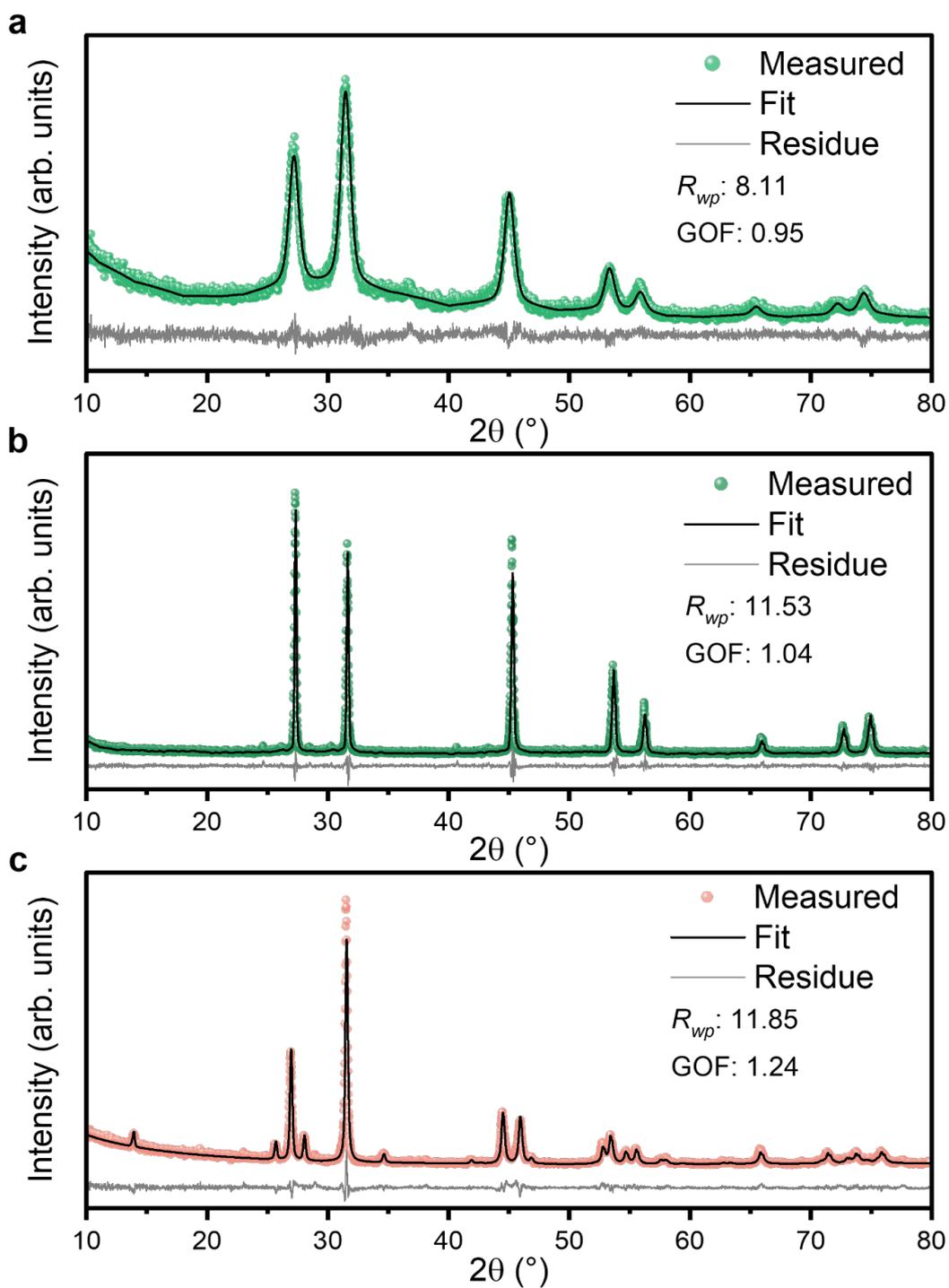

**Supplementary Fig. 1 | Rietveld refinement of the XRD patterns shown in Fig. 1c.** The measured data points (dots), fit (black lines) and residuals (grey lines) along with $R_{wp}$ and goodness of fit are given for **a** cation-disordered cubic NCs, **b** cubic powders and **c** cation-ordered powders. Cubic-phase samples are fit against the $Fm\overline{3}m$ structure, while the cation-ordered sample was here fit with the $R\overline{3}m$ structure.

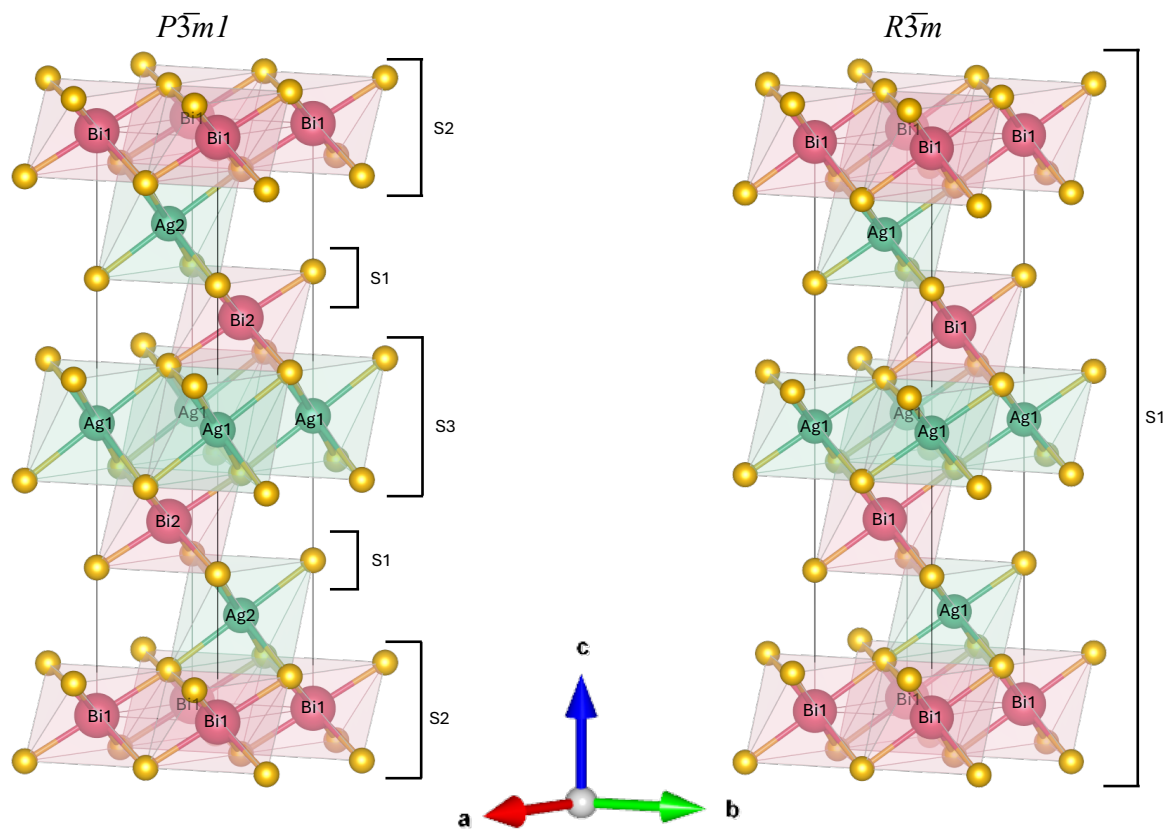

**Supplementary Fig. 2 | Unit cells of the *P3̄m1* and *R3̄m* phases of AgBiS$_2$** with the different atomic sites labelled. These were some of the initial proposed models as seen in the work carried out by Geller *et al[1]*. It was thought that the selenide and sulfide analogues may behave similarly, and that a low temperature *P3̄m*1 phase would be seen for AgBiS$_2$.

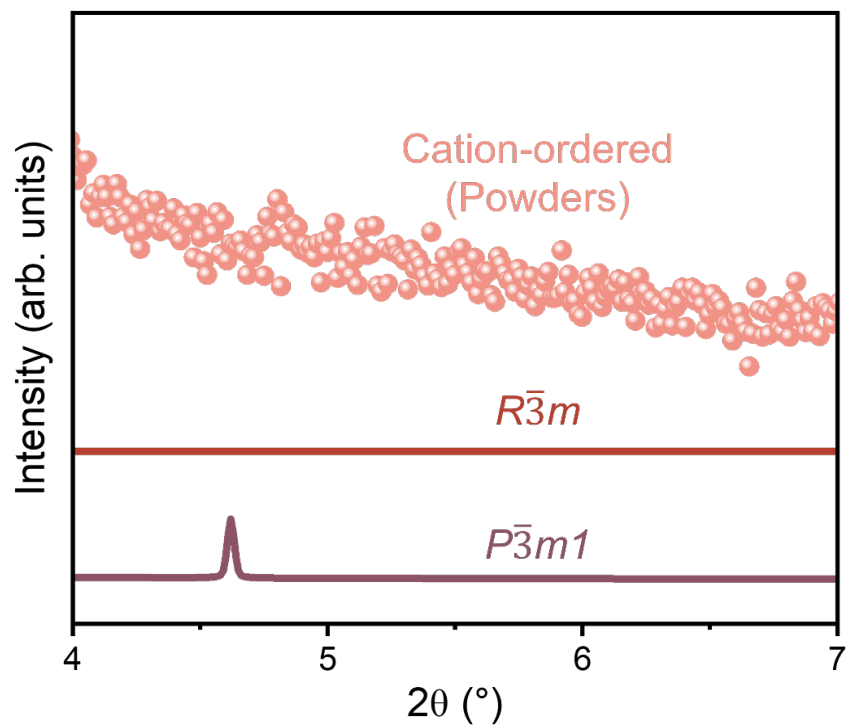

**Supplementary Fig. 3 | Zoom-in of the XRD patterns of cation-ordered powders** showing there to be no peak at around 4.63°. This shows that these powder samples are not consistent with the $P\bar{3}m1$ phase.

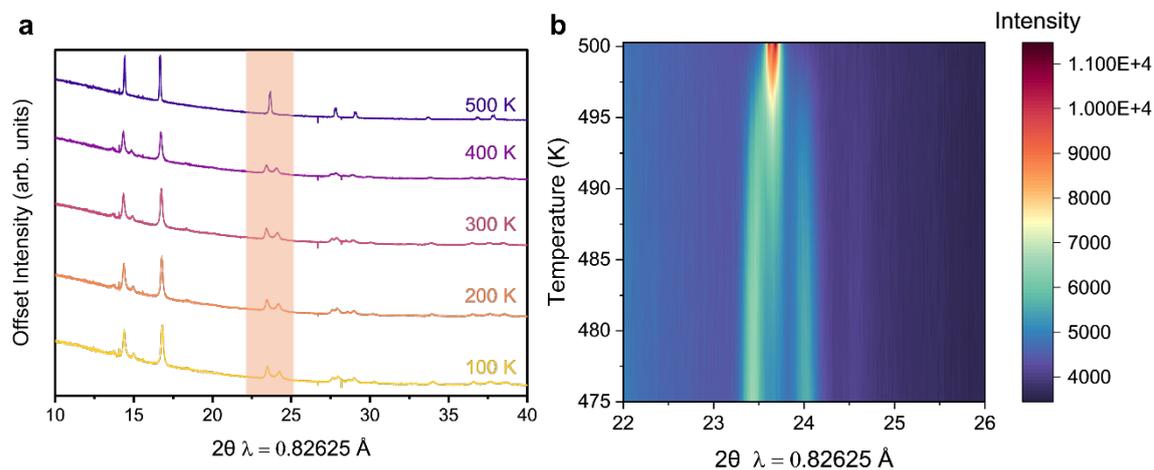

**Supplementary Fig. 4 | Phase transitions for cation-ordered AgBiS₂. a,** Waterfall plot and **b,** contour plot of PXRD patterns showing the phase change of the sample from cation-ordered to the cation-disordered cubic form as it was heated *in-situ*. The evolution of the peak shapes from the cation-ordered to cation-disordered cubic phase can be seen clearly in the waterfall plot. These measurements were carried out at room temperature at the Diamond Light Source on the I11 Beamline. The sample was packed into a 0.5 mm borosilicate glass capillary with ground glass added to the sample to reduce preferred orientation and absorption effects.

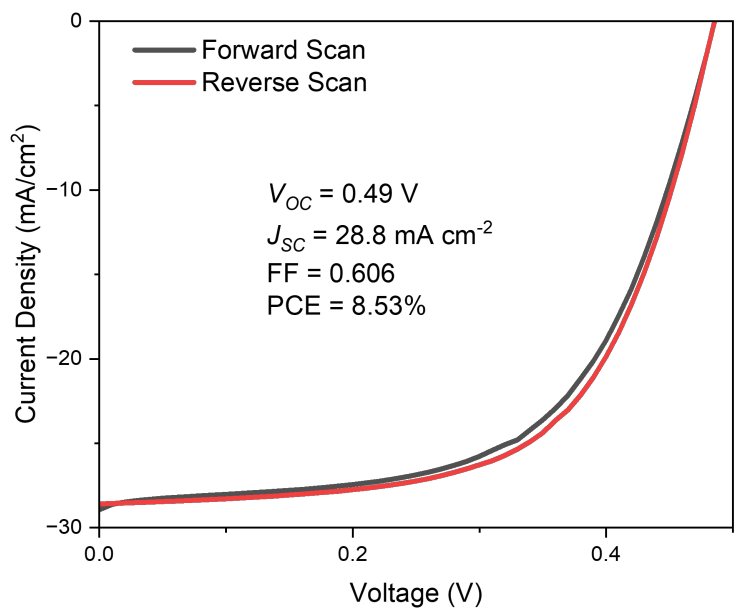

**Supplementary Fig. 5 | *J–V* curves of a photovoltaic device made from colloidal cubic AgBiS₂ NCs.** Solution-phase ligand-exchange was used, following prior reports[2]. The device structure was ITO/SnO$_2$/AgBiS$_2$/AgBiS$_2$-EDT/PTAA/MoO$_x$/Ag, and devices were tested under calibrated 1-sun illumination in air.

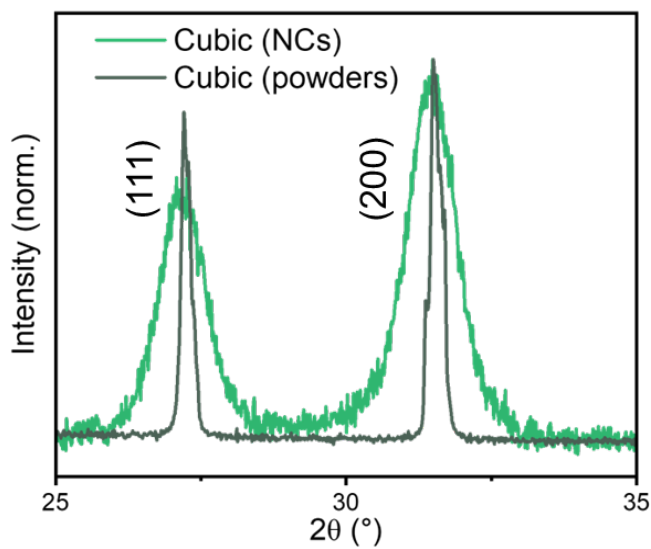

**Supplementary Fig. 6 | Zoom-in of the XRD patterns of cubic powders compared with cubic AgBiS₂ NCs** post-treated at 110 °C. The intensities of the XRD patterns are normalized.

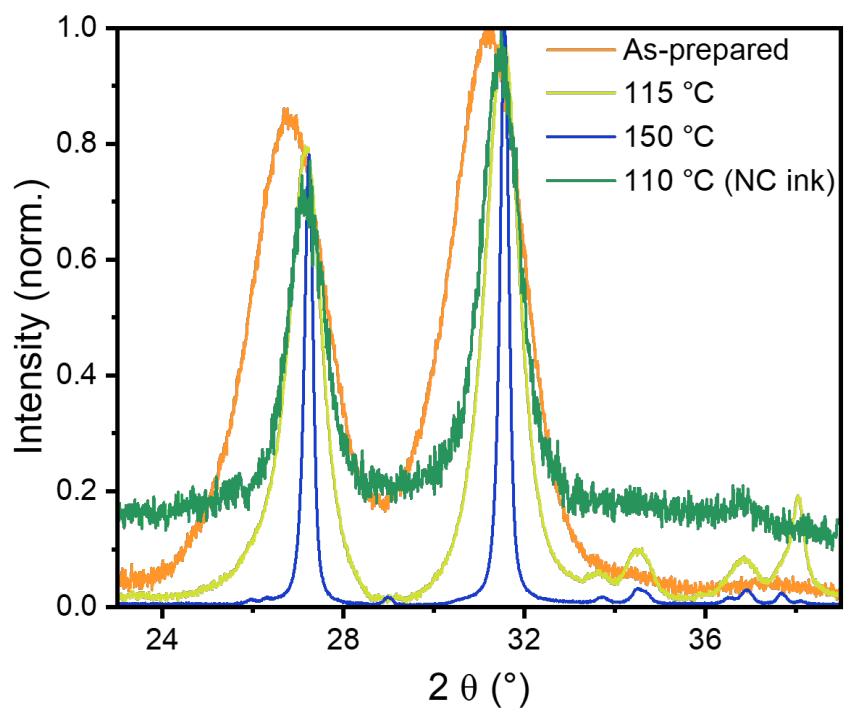

**Supplementary Fig. 7 | Room-temperature XRD pattern** of cubic nanocrystal (NC) ink post-treated at 110 °C, compared with XRD patterns of cubic NCs reported in Ref. 3 ( labelled "as-prepared", "115 °C" and "150 °C").

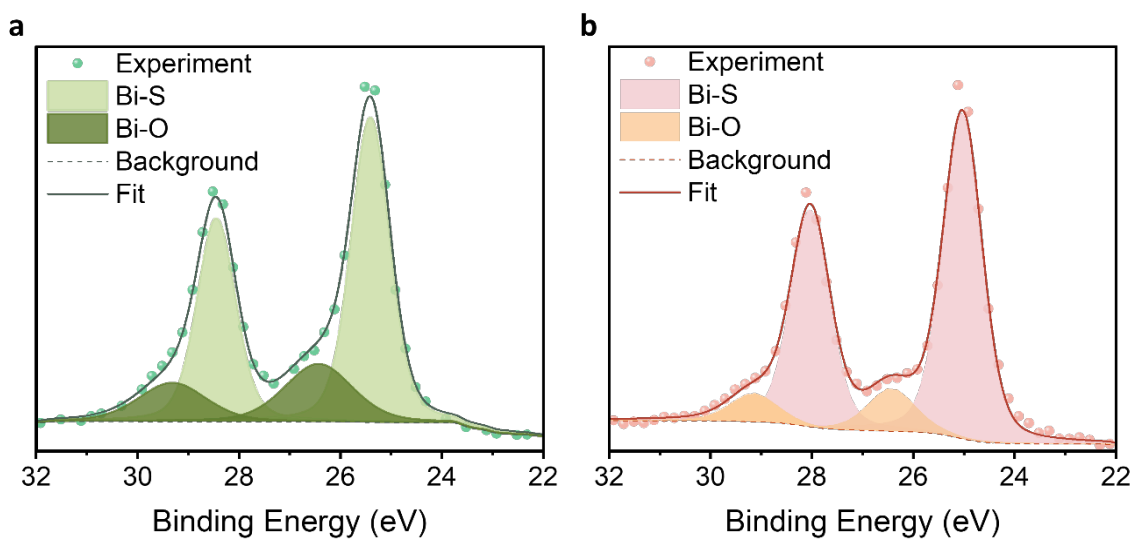

**Supplementary Fig. 8 | XPS Bi 5*d* core level spectra** fit using a Shirley background and Gaussian-Lorentzian sum (SGL) peak shape, showing Bi-S as well as surface Bi-O peaks for **a** cubic NCs and **b** cation-ordered powders.

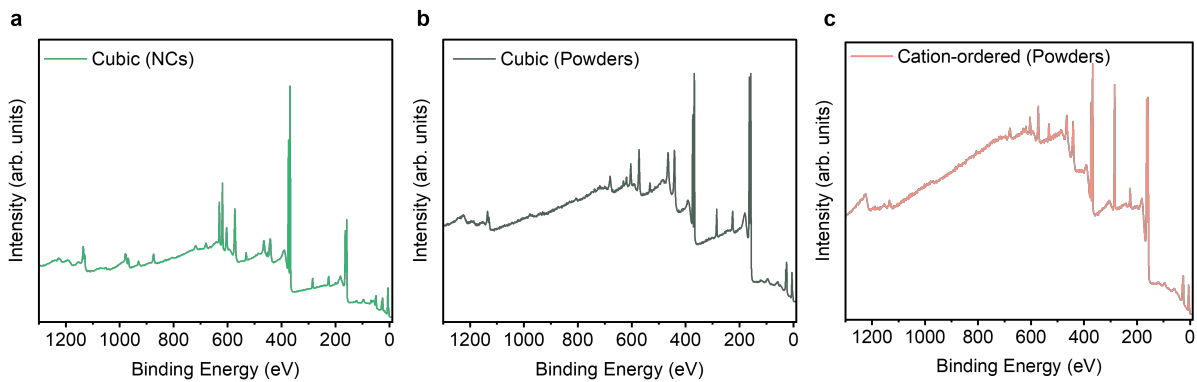

**Supplementary Fig. 9 | XPS survey spectra of AgBiS₂ samples.** Measurements for **a** cubic NCs, **b** cubic powders and **c** cation-ordered powders.

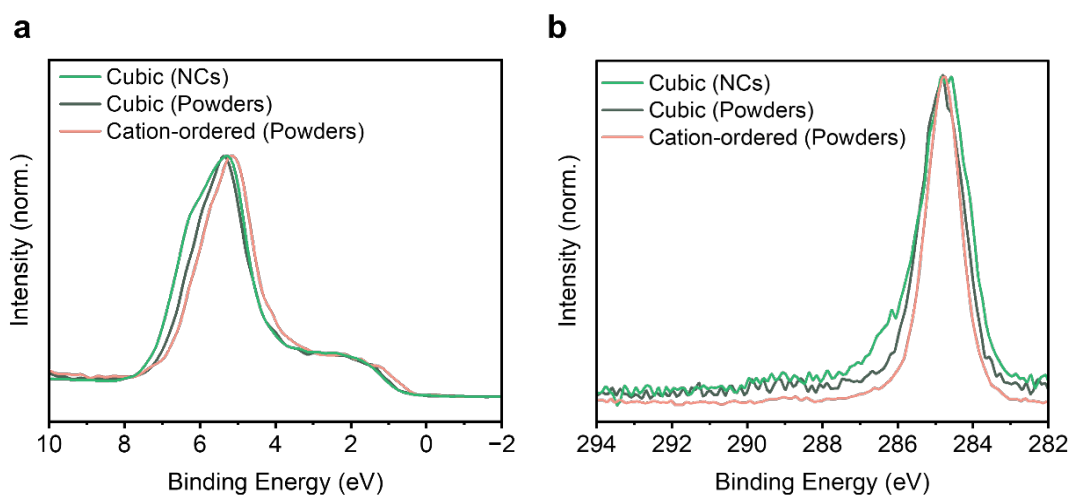

**Supplementary Fig. 10 | XPS spectra of cubic NCs, cubic powders, and cation-ordered powders of AgBiS₂** at the **a** leading edge, and **b** C 1*s* core level.

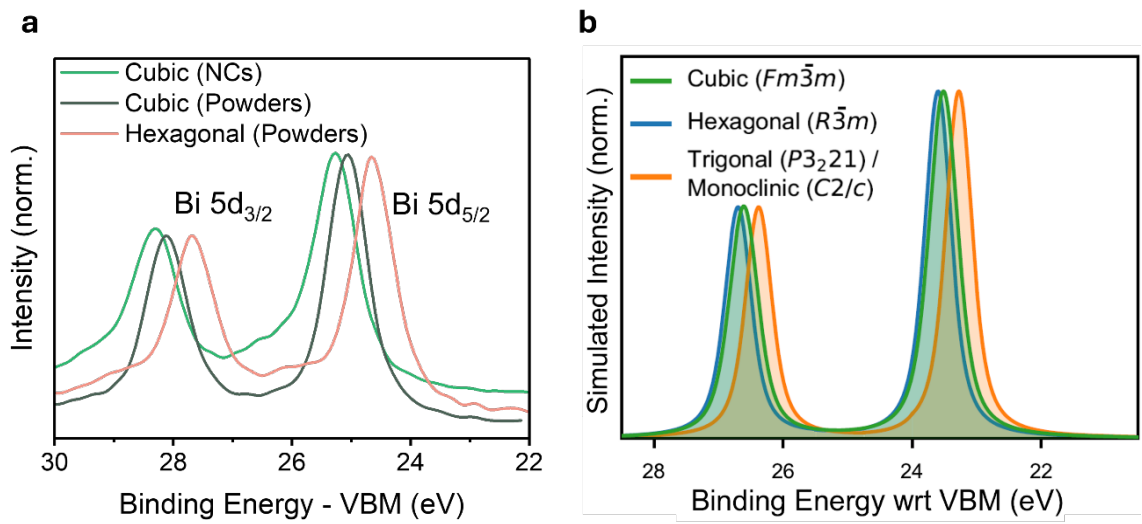

**Supplementary Fig. 11 | Comparison between measured and calculated XPS core level spectra. a** Bi 5d core levels measured from XPS of cubic powders and nanocrystals, and cation-ordered powders of AgBiS$_2$. The binding energy scale was shifted to be relative to the valence band maximum (VBM) in order to directly compare with the computations. **b** Calculated XPS spectra (as in Fig. 1d).

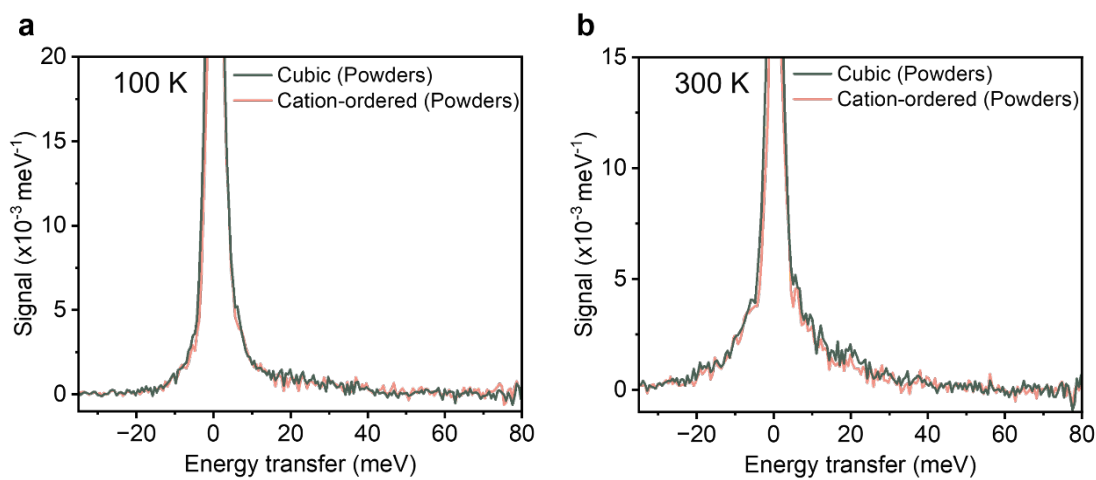

**Supplementary Fig. 12 | Inelastic neutron scattering (INS) spectra** of the cubic and cation-ordered AgBiS$_2$ powders at **a** 100 K and **b** 300 K. The INS spectra of the two phases are almost indistinguishable at both 100 K and 300 K.

**Supplementary Note 2 | Detailed structural analysis of cation-ordered AgBiS₂**

The computed $C2/c$ and experimentally-determined $P3_221$ structures are similar to the $R\bar{3}m$ phase, with the same layering of Ag-S and Bi-S octahedra, but now with the Ag(I) and Bi(III) cations off-centre, and the AgS₆ and BiS₆ octahedra distorted. As shown in Supplementary Fig. **13**c, in going from the $R\bar{3}m$ to $C2/c$ phase, there are collective shifts of the sulfur atoms in the plane of the cation/anion layers, along with some minor movements of the Bi atoms. The S layer shifts are in-phase for those sandwiching the Bi layers, and out-of-phase across Ag layers (*i.e.*, following an AABB pattern across -S-Bi-S-Ag-S-Bi-S- layers). The AgS₆ octahedra distort more than the BiS₆ octahedra. As shown in Supplementary Fig. **13** and Supplementary Table 1-4, this distortion lowers the structural symmetry from a single cation-anion bond length per octahedron in the $R\bar{3}m$ phase, to three pairs of bond lengths for Ag-S (from ~2.5 – 3.1 Å) and Bi-S (from ~2.7 – 3.1 Å) in the $C2/c$ phase. The $P3_221$ phase is similar to the computed $C2/c$ structure, in that it also has Ag(I) and Bi(III) off centre.

To understand the three phases ($R\bar{3}m$, $C2/c$ and $P3_221$), we analyzed the distributions in bond lengths and angles (Supplementary Table 1-4). The distortions of the octahedra were quantified with the bond length distortion index ($D$) and bond angle variance ($\sigma(\theta)^2$), as detailed in Supplementary Eq. S1-S2[4, 5].

$$D = \frac{1}{6} \sum_{i=1}^{6} \frac{|d_i - d_0|}{d_0} \qquad (S1)$$

$$\sigma(\theta)^2 = \frac{1}{12} \sum_{i=1}^{12} (\theta_i - 90)^2 \qquad (S2)$$

where $d_i$ and $d_0$ represent each cation-S bond length and mean bond length, respectively, while $\theta_i$ corresponds to S-cation-S bond angles. $D$ and $\sigma(\theta)^2$ are zero for a fully-symmetric octahedron, while larger $D$ and $\sigma(\theta)^2$ values imply a more distorted octahedron. Note that many prior literature would calculate $\sigma(\theta)^2$ by dividing by 11 rather than 12; we here divide by 12, since we are specifically comparing against an ideal octahedron that has 12 bond angles. The values for all RPA@PBE computationally-relaxed phases are shown in Table 3. For a perfectly cubic $Fm\overline{3}m$ phase, both $D$ and $\sigma(\theta)^2$ would be zero, however, local octahedral distortions in the disordered material result in non-zero values in reality. The $R\overline{3}m$ phase has a single bond length within each octahedron (giving zero values for $D$), but the bond angles deviate from 90°, with $AgS_6$ octahedra exhibiting a larger deviation. By contrast, both $C2/c$ and $P3_221$ phases exhibit larger $D$ and $\sigma(\theta)^2$ values, particularly for the $AgS_6$ octahedra. This analysis of bond lengths and angles shows that the $AgS_6$ octahedra distort more than $BiS_6$, which contrasts to prior assumptions for $AgBiS_2$, where it was assumed that distortions mostly originated from the off-centering of Bi(III)[6].

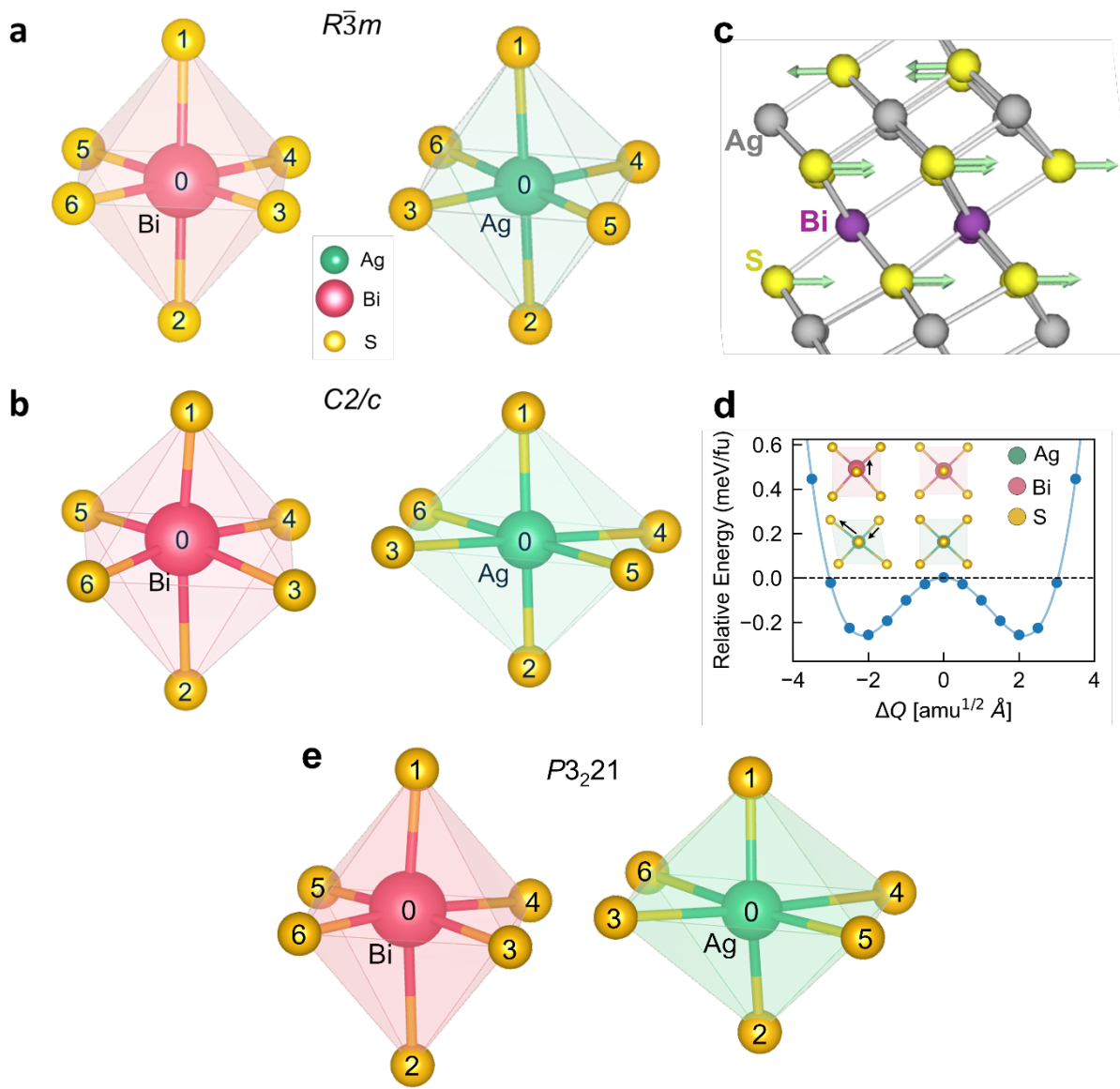

**Supplementary Fig. 13 | Illustration of AgS$_6$ and BiS$_6$ octahedra** in **a** $R\overline{3}m$, **b** $C2/c$ and **e** $P3_221$ AgBiS$_2$. The green, red, and yellow spheres refer to Ag, Bi, and S ions, respectively. Bond lengths and angles detailed in Supplementary Table 1-4. **c** Illustration of the imaginary phonon mode of $R\overline{3}m$ shown in the phonon dispersion curve in Fig. 2b. The imaginary phonon mode at the Z point in reciprocal space primarily involves shearing of the S atom positions. The grey, purple and yellow spheres represent Ag, Bi and S ions, respectively. **d** Configuration coordinate diagram along the path of the imaginary phonon mode of $R\overline{3}m$ at Z in Fig. 2b, calculated using semi-local Density Functional Theory (DFT) – PBEsol. $\Delta Q$ represents the mass-weighted atomic displacements of atoms from equilibrium positions in the $R\overline{3}m$ structure, and the energy is given per formula-unit. The two valleys near $\pm 2$ amu$^{1/2}$Å $Q$-values correspond to distortions that yield the $C2/c$ structure. Note that the final energy difference between $R\overline{3}m$ and $C2/c$ (~52 meV/formula-unit) when fully relaxing from this distorted geometry and using RPA is much larger than implied here.

**Supplementary Table 1** | Ag-S bond lengths for the $R\bar{3}m$, $C2/c$ and $P3_221$ phases, calculated using RPA@PBE.

| Ag-S bond | Ag-S bond length (Å) | | |
|:---:|:---:|:---:|:---:|
| | $R\bar{3}m$ | $C2/c$ | $P3_221$ |
| 0-1 | 2.776 | 2.473 | 2.479 |
| 0-2 | 2.776 | 2.473 | 2.479 |
| 0-3 | 2.776 | 3.008 | 2.828 |
| 0-4 | 2.776 | 3.008 | 3.402 |
| 0-5 | 2.776 | 3.104 | 2.828 |
| 0-6 | 2.776 | 3.104 | 3.402 |

**Supplementary Table 2** | S-Ag-S bond angles for the $R\bar{3}m$, $C2/c$ and $P3_221$ phases, calculated using RPA@PBE.

| S-Ag-S bond | S-Ag-S bond angle (°) | | |
|:---:|:---:|:---:|:---:|
| | $R\bar{3}m$ | $C2/c$ | $P3_221$ |
| 1-0-3 | 92.3 | 94.5 | 91.6 |
| 1-0-4 | 87.7 | 85.5 | 86.0 |
| 1-0-5 | 87.7 | 85.8 | 99.9 |
| 1-0-6 | 92.3 | 94.2 | 79.1 |
| 2-0-3 | 87.7 | 85.5 | 99.9 |
| 2-0-4 | 92.3 | 94.5 | 79.1 |
| 2-0-5 | 92.3 | 94.2 | 91.6 |
| 2-0-6 | 87.7 | 85.8 | 86.0 |
| 3-0-5 | 87.7 | 97.3 | 111.4 |
| 3-0-6 | 92.3 | 82.7 | 80.9 |
| 4-0-5 | 92.3 | 82.7 | 80.9 |
| 4-0-6 | 87.7 | 97.3 | 86.7 |

**Supplementary Table 3** | Bi-S bond lengths for the $R\bar{3}m$, $C2/c$ and $P3_221$ phases, calculated using RPA@PBE.

| Bi-S bond | Bi-S bond length (Å) | | |
|---|---|---|---|
| | $R\bar{3}m$ | $C2/c$ | $P3_221$ |
| 0-1 | 2.828 | 3.053 | 2.677 |
| 0-2 | 2.828 | 2.672 | 3.035 |
| 0-3 | 2.828 | 3.053 | 2.839 |
| 0-4 | 2.828 | 2.840 | 3.035 |
| 0-5 | 2.828 | 2.672 | 2.839 |
| 0-6 | 2.828 | 2.840 | 2.677 |

**Supplementary Table 4** | S-Bi-S bond angles for the $R\bar{3}m$, $C2/c$ and $P3_221$ phases, calculated using RPA@PBE.

| S-Bi-S bond | S-Bi-S bond angle (°) | | |
|---|---|---|---|
| | $R\bar{3}m$ | $C2/c$ | $P3_221$ |
| 1-0-3 | 89.9 | 81.1 | 95.0 |
| 1-0-4 | 89.9 | 85.4 | 90.6 |
| 1-0-5 | 90.1 | 91.4 | 91.8 |
| 1-0-6 | 90.1 | 86.5 | 96.5 |
| 2-0-3 | 90.1 | 91.4 | 84.8 |
| 2-0-4 | 90.1 | 94.2 | 82.3 |
| 2-0-5 | 89.9 | 96.0 | 87.6 |
| 2-0-6 | 89.9 | 93.0 | 90.6 |
| 3-0-4 | 90.1 | 86.5 | 87.6 |
| 3-0-6 | 89.9 | 85.4 | 91.8 |
| 4-0-5 | 89.9 | 93.0 | 84.8 |
| 5-0-6 | 90.1 | 96.0 | 95.0 |

**Supplementary Table 5** | Relative energies in meV/atom of hexagonal ($R\overline{3}m$, $P3m1$[7]), monoclinic ($C2/c$) and trigonal ($P3_221$) AgBiS$_2$ phases, relaxed and calculated using various levels of theory. The Random Phase Approximation (RPA@PBE) is considered the most accurate level of theory here. When including the D3 dispersion correction[8] with semi-local DFT (PBE, PBEsol), the $C2/c$ and $P3_221$ phases relaxed to the $R\overline{3}m$ structure, and so no energy is reported for these cases.

| Energy Functional | $R\overline{3}m$ | $P3m1$ | $C2/c$ | $P3_221$ |
|---|---|---|---|---|
| PBE | 0 | -22.8 | -16.7 | -17.7 |
| PBE + D3 | 0 | +25.3 | – | – |
| PBEsol | 0 | +1.0 | -0.5 | -2.8 |
| PBEsol + SOC | 0 | +6.0 | -0.4 | -1.8 |
| PBEsol + D3 | 0 | +43.9 | – | – |
| HSE06 | 0 | -22.7 | -18.7 | -19.7 |
| HSE06 + D3 | 0 | +29.1 | -0.3 | -0.2 |
| PBE0 | 0 | -24.4 | -21.5 | -22.3 |
| PBE0 + D3 | 0 | +23.2 | -0.6 | -3.3 |
| RPA@PBE | 0 | +9.9 | -12.8 | -13.3 |

As shown in Supplementary Table 5, we note that the relative energies of the various ordered crystal polymorphs of AgBiS$_2$ are remarkably sensitive to the choice of Density Functional Theory (DFT-) exchange-correlation functional and inclusion of dispersion corrections. Strong functional dependence in the Ag-S bond lengths for AgBiS$_2$ was noted in our previous work[3], which was attributed to the localisation of the semi-valence Ag 4$d$ orbitals. This functional dependence will similarly affect predictions of the order-disorder transition temperature in this system.

**Supplementary Table 6** | Equilibrium lattice parameters and volume per atom for hexagonal ($R\bar{3}m$) AgBiS$_2$, calculated using various levels of theory.

| Energy Functional | $a$ (Å) | $\alpha$ (°) | Volume (Å$^3$/atom) |
|:---:|:---:|:---:|:---:|
| PBE | 6.77 | 34.74 | 22.53 |
| PBE + D3 | 6.62 | 35.21 | 21.50 |
| PBEsol | 6.58 | 35.32 | 21.22 |
| PBEsol + D3 | 6.45 | 35.71 | 20.41 |
| HSE06 | 6.79 | 34.39 | 22.28 |
| HSE06 + D3 | 6.59 | 35.00 | 21.06 |
| PBE0 | 6.79 | 34.42 | 22.27 |
| PBE0 + D3 | 6.61 | 34.95 | 21.20 |
| RPA@PBE | 6.74 | 34.54 | 21.97 |

**Supplementary Table 7** | Equilibrium lattice parameters and volume per atom for hexagonal ($P3m1$) AgBiS$_2$, calculated using various levels of theory.

| Energy Functional | $a$ (Å) | $c$ (Å) | Volume (Å$^3$/atom) |
|:---:|:---:|:---:|:---:|
| PBE | 4.11 | 6.86 | 25.07 |
| PBE + D3 | 4.05 | 6.78 | 24.05 |
| PBEsol | 4.05 | 6.69 | 23.76 |
| PBEsol + D3 | 4.00 | 6.62 | 22.95 |
| HSE06 | 4.07 | 6.87 | 24.64 |
| HSE06 + D3 | 4.00 | 6.77 | 23.47 |
| PBE0 | 4.07 | 6.87 | 24.62 |
| PBE0 + D3 | 4.01 | 6.78 | 23.57 |
| RPA@PBE | 4.04 | 6.82 | 24.04 |

**Supplementary Table 8 |** Equilibrium lattice parameters and volume per atom for monoclinic ($C2/c$) AgBiS$_2$, calculated using various levels of theory. When including the D3 dispersion correction[8] with semi-local DFT (PBE, PBEsol), the $C2/c$ phase relaxed to the $R\overline{3}m$ structure, and so no lattice parameters are reported for these cases.

| Energy Functional | $a$ (Å) | $c$ (Å) | Volume (Å$^3$/atom) |
|---|---|---|---|
| PBE | 4.13 | 13.14 | 24.56 |
| PBE + D3 | – | – | – |
| PBEsol | 4.00 | 12.53 | 21.49 |
| PBEsol + D3 | – | – | – |
| HSE06 | 4.09 | 13.16 | 24.20 |
| HSE06 + D3 | 3.97 | 12.58 | 21.06 |
| PBE0 | 4.09 | 13.15 | 24.21 |
| PBE0 + D3 | 3.97 | 12.62 | 21.20 |
| RPA@PBE | 4.04 | 12.92 | 22.92 |

**Supplementary Table 9 |** Equilibrium lattice parameters and volume per atom for trigonal ($P3_221$) AgBiS$_2$, calculated using various levels of theory. When including the D3 dispersion correction[8] with semi-local DFT (PBE, PBEsol), the $P3_221$ phase relaxed to the $R\overline{3}m$ structure, and so no lattice parameters are reported for these cases.

| Energy Functional | $a$ (Å) | $c$ (Å) | Volume (Å$^3$/atom) |
|---|---|---|---|
| PBE | 4.13 | 19.40 | 23.88 |
| PBE + D3 | – | – | – |
| PBEsol | 4.04 | 18.68 | 22.01 |
| PBEsol + D3 | – | – | – |
| HSE06 | 4.09 | 19.40 | 23.44 |
| HSE06 + D3 | 4.00 | 18.65 | 21.52 |
| PBE0 | 4.09 | 19.41 | 23.48 |
| PBE0 + D3 | 4.02 | 18.75 | 21.82 |
| RPA@PBE | 4.07 | 19.28 | 23.02 |

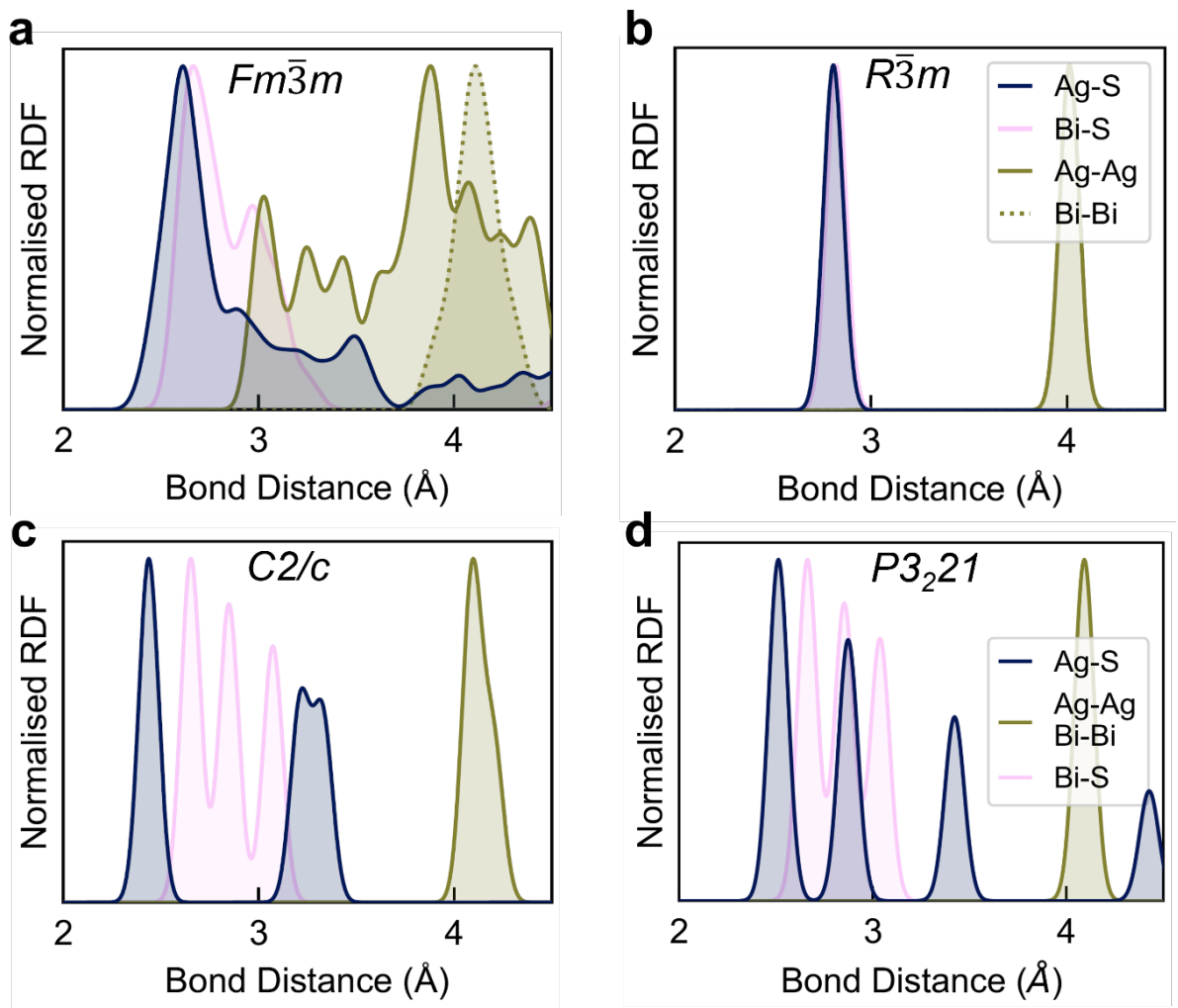

**Supplementary Fig. 14 | Calculated radial distribution function (RDF) for cation-anion and cation-cation bonds** in the **a** disordered cubic $Fm\bar{3}m$, **b** hexagonal $R\bar{3}m$, **c** monoclinic $C2/c$ and **d** trigonal $P3_221$ AgBiS$_2$ structures. Gaussian smearing of width 0.05 Å was applied to mimic thermal broadening effects.

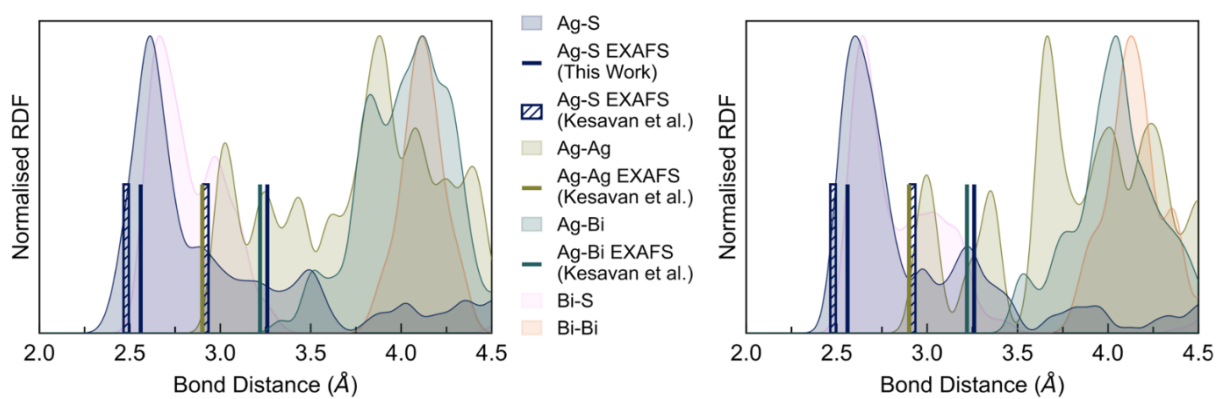

**Supplementary Fig. 15 | Calculated radial distribution function (RDF) for cation-anion and cation-cation bonds** in disordered cubic $Fm\bar{3}m$ AgBiS$_2$, using 128-atom (left) and 64-atom (right) SQS supercells with

HSE06 hybrid DFT. Gaussian smearing of width 0.05 Å was applied to mimic thermal broadening effects, and bond lengths extracted from experimental EXAFS measurements[9] are marked in bars.

As shown in Supplementary Fig. 15, the Ag-S RDF in disordered $AgBiS_2$ peaks around 2.5-2.6 Å, but shows a wide tail up to ~3.5 Å, matching EXAFS measurements of multiple Ag-S bond lengths in cation-disordered $AgBiS_2$. This distribution of Ag-S bond lengths is the result of similar off-centring and distortion of Ag-S octahedra witnessed for the $P3_221$ ordered phase. Moreover, we find that in regions of higher Ag density (due to local fluctuations in the random cation distribution), local Ag-S bonding approaches that of $Ag_2S$, with strong octahedral distortions and short Ag-Ag distances, as observed in EXAFS measurements from Kesavan et al[9]. We note that SQS supercells assume random site disorder, neglecting potential short-range ordering effects which could also play a role in the heterovalent cation distributions.

In addition to fitting with the $P3_221$ phase in the main text, we also here consider fitting the Ag K-edge EXAFS with the $C2/c$ model, from which two Ag–S distances are extracted (Supplementary Table 10). The $\sigma^2$ of cation-ordered Ag–S distance at 3.33 Å is more than double that of the shorter one, indicating significantly higher local disorder/distance variation for the S ions that are further away, or the presence of two inequivalent bonds of similar length. Since both fittings against $C2/c$ and $P3_221$ models demonstrate good fit, we cannot distinguish between these two phases based on EXAFS results. Hence, further high-resolution structural analysis is required (as detailed in the main text).

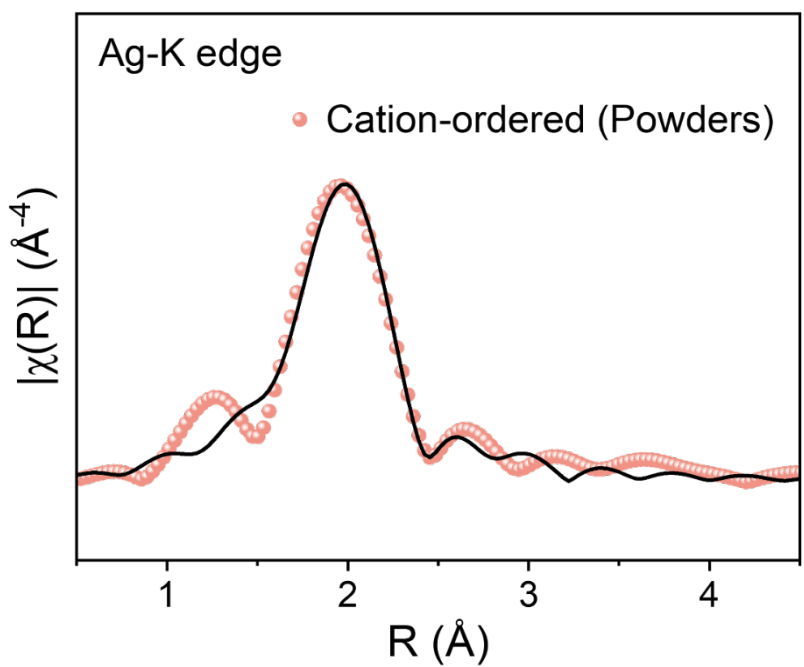

**Supplementary Fig. 16** | $k^2$-weighted transmission Ag K-edge FT-EXAFS moduli (phase uncorrected) of the cation-ordered AgBiS$_2$ powders. The EXAFS fitting was performed against the *C2/c* model. Corresponding bond lengths given in Supplementary Table 10. The main text shows fitting to the *P*3$_2$21 model.

**Supplementary Table 10** | Bond length ($R$) along with Debye–Waller factor ($\sigma^2$) for different Ag-S scattering paths extracted from the fittings to the Ag K-edge EXAFS spectra of cation-ordered AgBiS$_2$ powders using the *C*2/*c* structural model.

| Sample | Scattering Path | $R$ (Å) | $\sigma^2$ (Å$^2$) |
|---|---|---|---|
| Cation-ordered | Ag-S$_1$ | 2.51(1) | 0.009(2) |
| | Ag-S$_2$ | 3.33(7) | 0.02(1) |

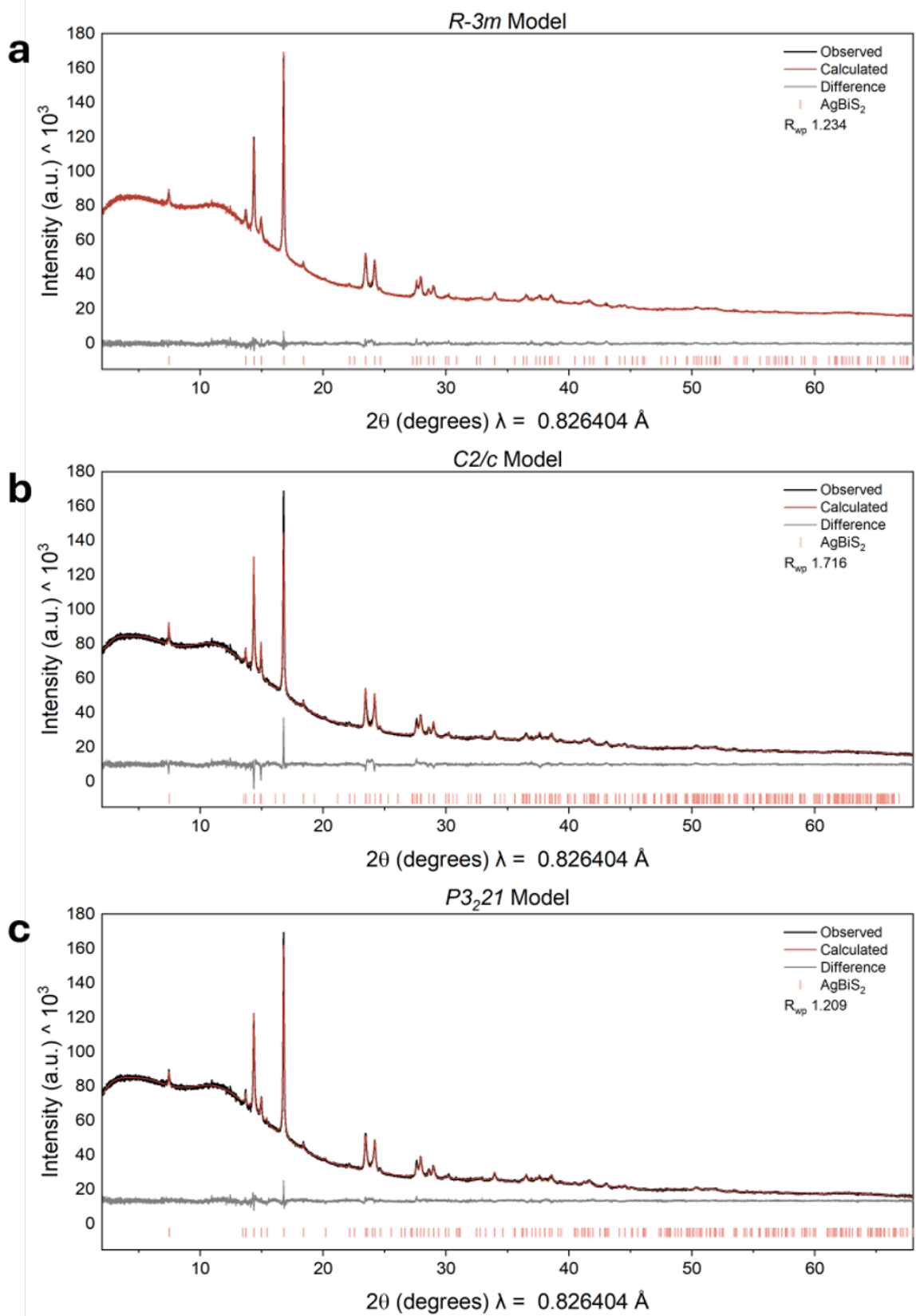

**Supplementary Fig. 17 | Comparative Rietveld plots of synchrotron PXRD data of the cation-ordered phase**, fitted to **a** $R\overline{3}m$, **b** $C2/c$ and **c** $P3_2 21$ space group models. The samples were highly absorbing, even when diluted more than half by ground glass, leading to low intensities being recorded.

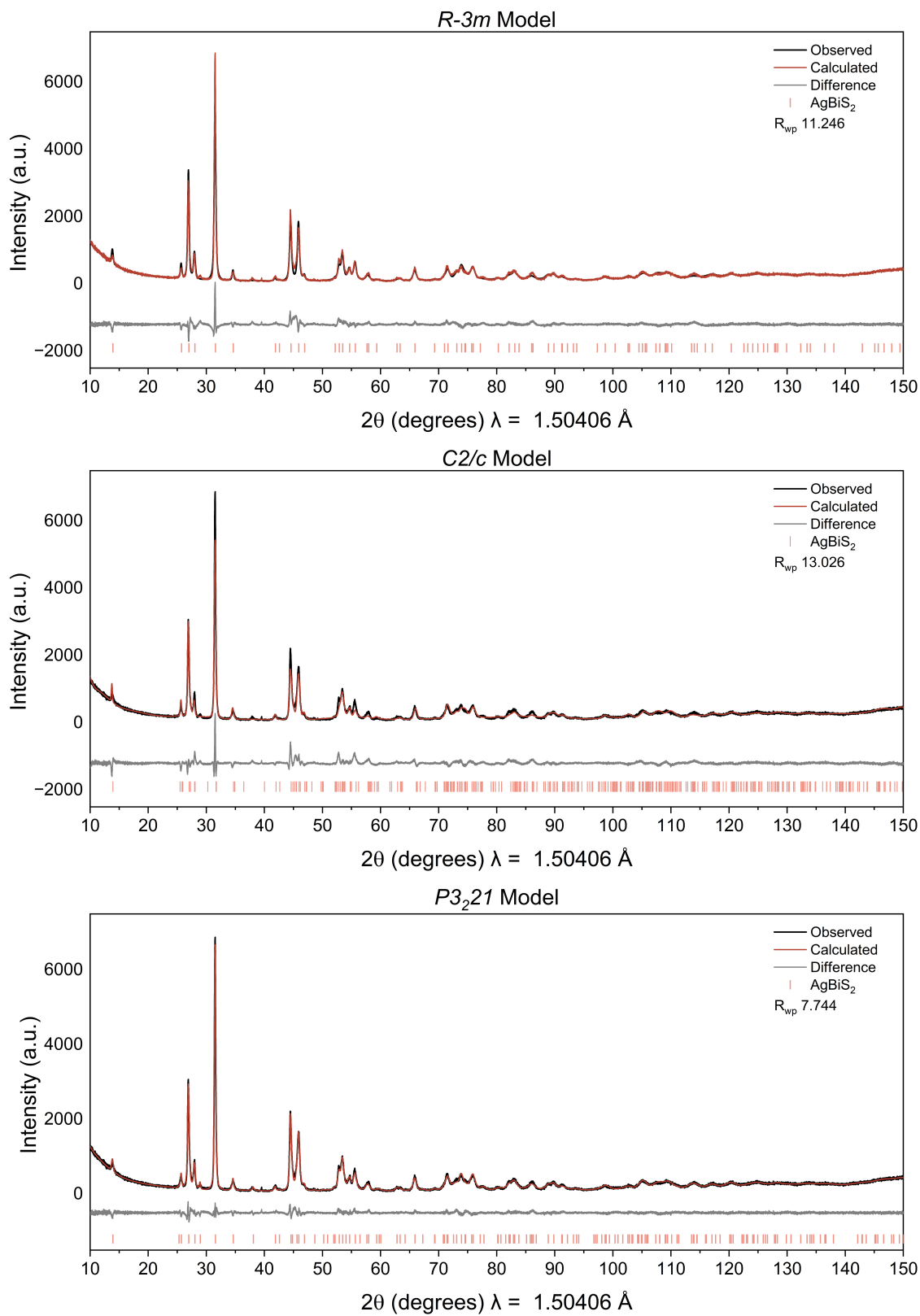

**Supplementary Fig. 18 | Comparative Rietveld plots of high-resolution laboratory PXRD data**, fitted to **a** $R\overline{3}m$, **b** $C2/c$ and **c** $P3_221$ space group models. Due to the issues surrounding the high background of the synchrotron XRD patterns in Supplementary Fig. 17, an overnight scan was performed on a Rigaku SmartLab

instrument here. A monochromator was used to ensure only the $K\alpha_1$ peaks were present. Multiple scans were performed and the results summed to allow for the visualisation of the lowest intensity peak.

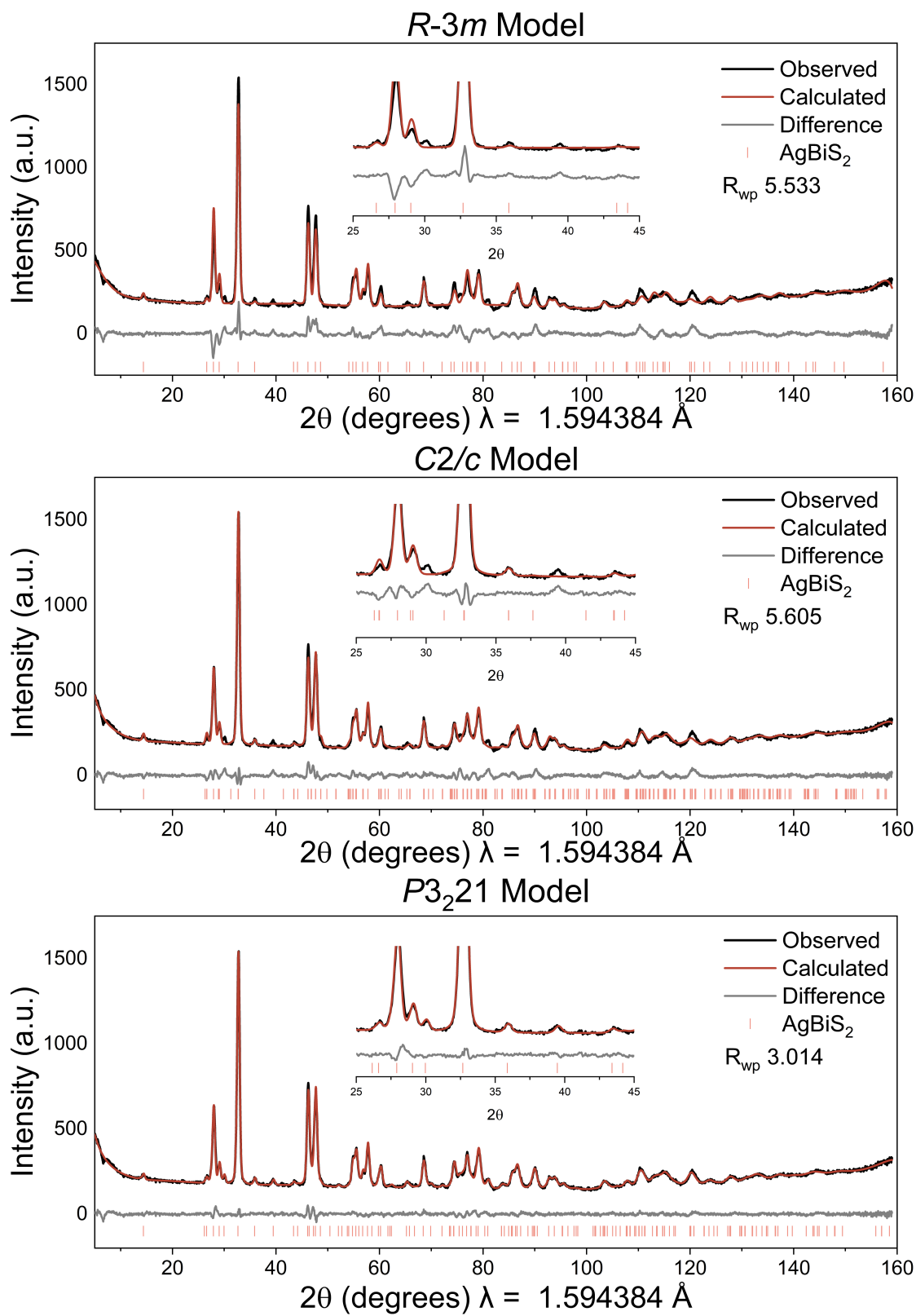

**Supplementary Fig. 19 | Comparative Rietveld plots of room temperature neutron data of the cation-ordered phase**, fitted to **a** $R\bar{3}m$, **b** $C2/c$ and **c** $P3_221$ space group models. Inset shows the region of $2\theta = 25\text{-}45°$ to more clearly illustrate how the different models account for different aspects of the data.

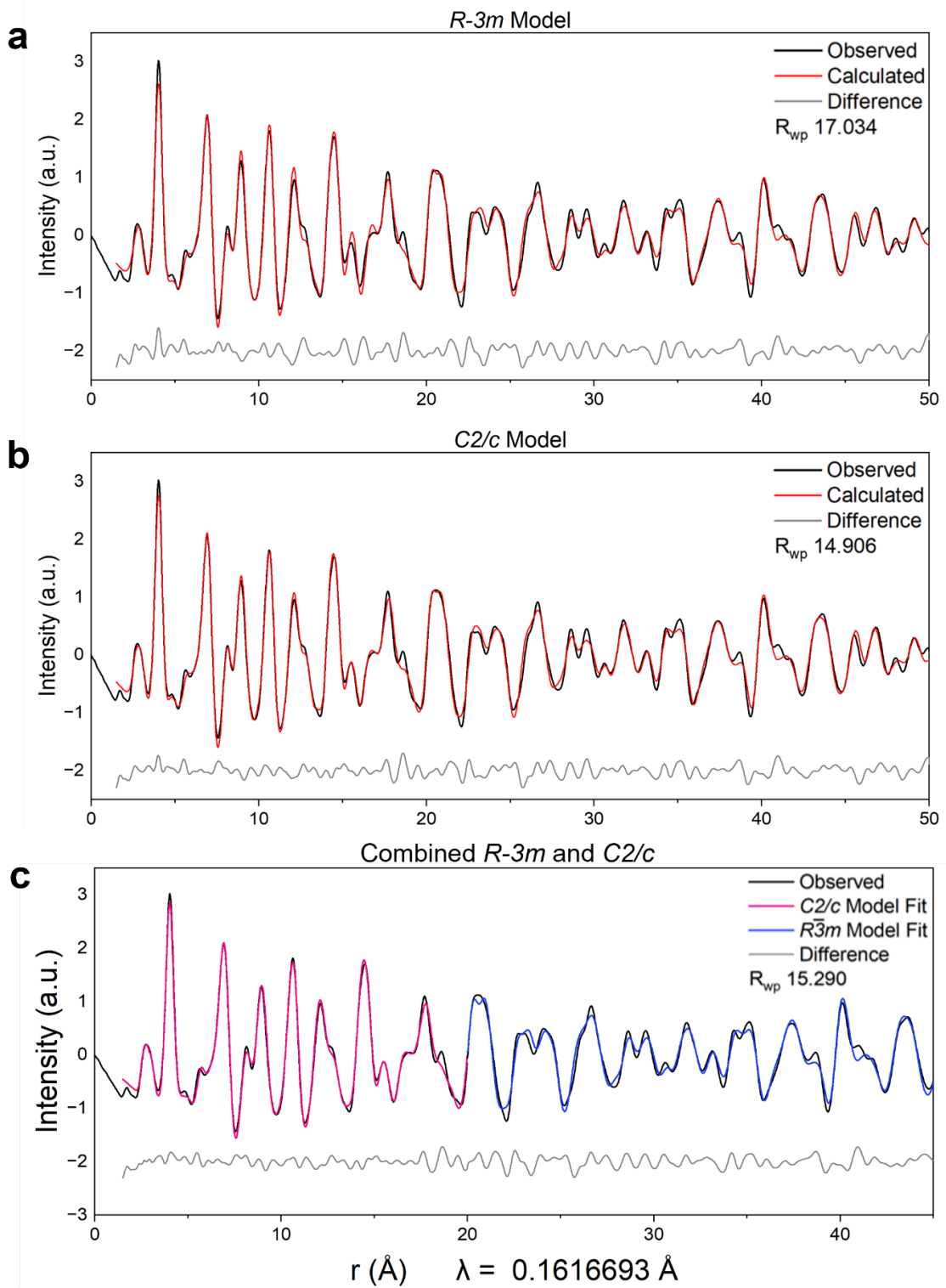

**Supplementary Fig. 20 | Comparative Rietveld plots of PDF data**, fitted to **a** $R\bar{3}m$ and **b** $C2/c$ space group models. **c** A model that fits $C2/c$ at low $r$ and $R\bar{3}m$ at high $r$ values, whilst allowing for the refinement of the point at which the model switches from one to the other. The fit of the $P3_221$ model to the PDF data can be seen in the main text Fig. 3f.

Attempts were made to fit the PDF data using the $C2/c$ model at low $r$ values, and the $R\overline{3}m$ at high $r$ values with the hypothesis that outside of the correlation length for a cooperative distortion, the symmetry-broken $C2/c$ model should no longer fit the data as effectively. The value for the correlation length was also allowed to refine and was found to be 20 Å. This was initially thought to indicate that small cooperative regions of $C2/c$-like structures exist within the sample, each with distortions in different directions to give a sum $R\overline{3}m$ structure macroscopically. However, the fit of the model to the $P3_221$ model is still superior to this combined model.

**Supplementary Table 11** | Fitting statistics for Supplementary Fig. 17-20.

| | Space Group Model | $R_{wp}$ | GOF |
|---|---|---|---|
| | $R\overline{3}m$ | 6.0136 | 14.124 |
| RT Neutron Refinements | $C2/c$ | 5.301 | 15.156 |
| | $P3_221$ | 4.076 | 11.653 |
| | $R\overline{3}m$ | 10.523 | 22.573 |
| 2 K Neutron Refinements | $C2/c$ | 5.528 | 11.863 |
| | $P3_221$ | 4.380 | 9.400 |
| | $R\overline{3}m$ | 17.034 | 0.109 |
| PDF Refinements | $C2/c$ | 14.910 | 0.096 |
| | $P3_221$ | 13.522 | 0.087 |
| | $R\overline{3}m$ | 1.234 | 3.203 |
| Synchrotron Refinements | $C2/c$ | 1.716 | 4.450 |
| | $P3_221$ | 1.209 | 3.136 |
| | $R\overline{3}m$ | 11.246 | 1.947 |
| Rigaku Laboratory Diffractometer Refinements | $C2/c$ | 13.027 | 2.257 |
| | $P3_221$ | 7.744 | 1.341 |

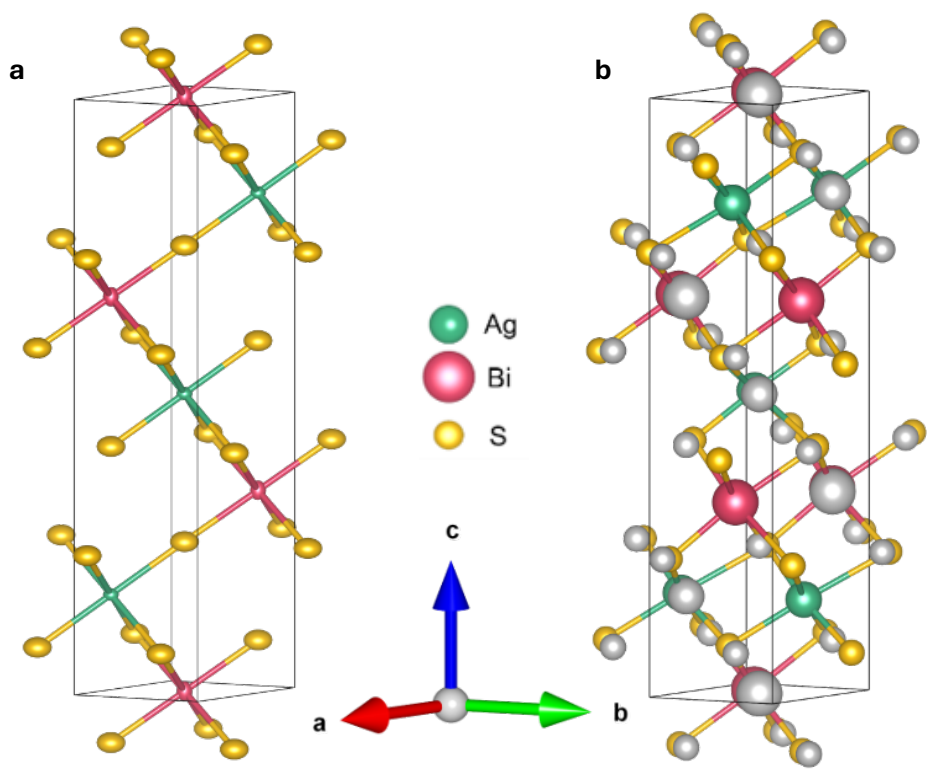

**Supplementary Fig. 21** | **Comparison of $R\overline{3}m$ and $P3_221$ model for cation-ordered AgBiS$_2$. a** The refined values for thermal ellipsoids of the $R\overline{3}m$ model from X-ray synchrotron measurements. The ellipsoids of the S anions are much larger than expected, suggesting that there are some problems with the model. **b** A comparison of the atom positions in $P3_221$ (in colour) to $R\overline{3}m$ (in grey), which shows that, while all atoms are displaced from their idealised $R\overline{3}m$ positions in the $P3_221$ model, the largest displacements are of the sulfur atoms. This would explain the large thermal ellipsoids of the sulfur.

**Supplementary Table 12 | Analysis of cation order in the trigonal phase of AgBiS$_2$.** An additional site was created for each of the metal ions and the cation occupancies of each metal site were allowed to refine with the constraint that the total occupancy of each metal must sum to 1. The results suggest that 95-100% of the metal ions are on the expected site.

| | RT Neutron | 2 K Neutron | PDF | Synchrotron | Rigaku |
|---|---|---|---|---|---|
| Ag occupancy of *3b* site | 0.946 ± 0.012 | 0.968 ± 0.016 | 0.996 ± 0.015 | 1.000 ± 0.007 | 0.925 ± 0.004 |
| Ag occupancy of *3a* site | 0.054 ± 0.012 | 0.032 ± 0.016 | 0.004 ± 0.015 | 0.000 ± 0.007 | 0.075 ± 0.004 |
| Bi occupancy of *3b* site | 0.054 ± 0.012 | 0.946 ± 0.016 | 0.004 ± 0.015 | 0.000 ± 0.007 | 0.075 ± 0.004 |
| Bi occupancy of *3a* site | 0.946 ± 0.012 | 0.032 ± 0.016 | 0.996 ± 0.015 | 1.000 ± 0.007 | 0.925 ± 0.004 |
| $R_{wp}$ | 4.094 | 4.447 | 13.521 | 1.415 | 10.242 |
| Goodness of Fit | 11.706 | 9.544 | 0.0871 | 4.626 | 1.773 |

**Supplementary Table 13 | Bi-S bond lengths from refinements of various investigations into the structure using the *P*3$_2$21 model with the bond lengths from the computed *P*3$_2$21 model for comparison. The atom numbering is shown in**

Supplementary Fig. **13**.

| Bi-S bond | Bond lengths (Å) | | | | | | |
|---|---|---|---|---|---|---|---|
| | RT Neutron | 2 K Neutron | X-ray PDF | Synchrotron XRD | Rigaku Laboratory XRD | DFT *P*3$_2$21 Model | Lissner *et al.*[10] |
| 0-1 | 3.038 | 3.072 | 3.083 | 3.055 | 2.965 | 2.677 | 3.048 |
| 0-2 | 2.753 | 2.728 | 2.738 | 2.742 | 2.965 | 3.035 | 2.678 |
| 0-3 | 3.038 | 3.072 | 3.083 | 3.055 | 2.708 | 2.839 | 3.048 |
| 0-4 | 2.840 | 2.898 | 2.941 | 2.778 | 2.708 | 3.035 | 2.863 |
| 0-5 | 2.753 | 2.728 | 2.738 | 2.742 | 2.998 | 2.839 | 2.678 |
| 0-6 | 2.840 | 2.898 | 2.941 | 2.778 | 2.999 | 2.677 | 2.863 |

**Supplementary Table 14** | Ag-S bond lengths from refinements of various investigations into the structure using the $P3_221$ model. The atom numbering is shown in

Supplementary Fig. **13**.

| Ag-S bond | Bond lengths (Å) | | | | | | |
|---|---|---|---|---|---|---|---|
| | RT Neutron | 2 K Neutron | X-ray PDF | Synchrotron XRD | Rigaku Laboratory XRD | DFT $P3_221$ Model | Lissner *et al.*[10] |
| 0-1 | 2.518 | 2.491 | 2.509 | 2.493 | 2.564 | 2.479 | 2.530 |
| 0-2 | 2.518 | 2.491 | 2.509 | 2.493 | 2.564 | 2.479 | 2.530 |
| 0-3 | 2.878 | 2.764 | 2.772 | 2.922 | 2.732 | 2.828 | 2.855 |
| 0-4 | 2.878 | 2.764 | 2.772 | 2.922 | 2.732 | 3.402 | 2.855 |
| 0-5 | 3.060 | 3.066 | 3.111 | 3.084 | 3.112 | 2.828 | 3.127 |
| 0-6 | 3.060 | 3.066 | 3.111 | 3.084 | 3.112 | 3.402 | 3.127 |

**Supplementary Table 15** | S-Bi-S bond angles from refinements of various investigations into the structure using the $P3_221$ model.

| S-Bi-S bond | S-Bi-S bond angle (°) | | | | | |
|---|---|---|---|---|---|---|
| | RT Neutron | 2 K Neutron | X-ray PDF | Synchrotron XRD | Rigaku Laboratory XRD | DFT $P3_221$ Model |
| 1-0-3 | 92.1 | 94.2 | 94.6 | 91.2 | 94.0 | 95.0 |
| 1-0-4 | 86.4 | 87.3 | 87.8 | 85.0 | 88.25 | 90.6 |
| 1-0-5 | 87.6 | 85.5 | 85.1 | 88.2 | 85.9 | 91.8 |
| 1-0-6 | 93.4 | 92.2 | 91.6 | 94.8 | 91.35 | 96.5 |
| 2-0-3 | 93.4 | 92.2 | 91.6 | 94.8 | 91.35 | 84.8 |
| 2-0-4 | 87.6 | 85.5 | 85.1 | 88.2 | 85.9 | 82.3 |
| 2-0-5 | 86.4 | 87.3 | 87.8 | 85.0 | 88.25 | 87.6 |
| 2-0-6 | 92.1 | 94.2 | 94.6 | 91.2 | 94.0 | 90.6 |
| 3-0-4 | 89.2 | 88.5 | 88.8 | 88.9 | 90.6 | 87.6 |
| 3-0-6 | 96.5 | 98.5 | 98.2 | 97.4 | 95.1 | 91.8 |
| 4-0-5 | 85.1 | 84.6 | 84.3 | 84.8 | 83.6 | 84.8 |
| 5-0-6 | 89.2 | 88.5 | 88.8 | 88.9 | 90.6 | 95.0 |

**Supplementary Table 16** | S-Ag-S bond angles from refinements of various investigations into the structure using the $P3_221$ model.

| S-Ag-S bond | S-Ag-S bond angle (°) | | | | | |
|---|---|---|---|---|---|---|
| | RT Neutron | 2 K Neutron | X-ray PDF | Synchrotron XRD | Rigaku Laboratory XRD | DFT $P3_221$ Model |
| 1-0-3 | 86.4 | 85.7 | 86.4 | 86.6 | 87.8 | 91.6 |
| 1-0-4 | 97.7 | 100.9 | 101.0 | 97.0 | 100.1 | 86.0 |
| 1-0-5 | 82.5 | 79.5 | 79.3 | 83.2 | 80.0 | 99.9 |
| 1-0-6 | 93.1 | 93.1 | 92.3 | 93.0 | 90.9 | 79.1 |
| 2-0-3 | 97.7 | 100.9 | 101.0 | 97.0 | 100.1 | 99.9 |
| 2-0-4 | 86.4 | 85.8 | 86.4 | 86.6 | 87.8 | 79.1 |
| 2-0-5 | 93.1 | 93.1 | 92.3 | 93.0 | 90.9 | 91.6 |
| 2-0-6 | 82.5 | 79.5 | 79.3 | 83.2 | 80.0 | 86.0 |
| 3-0-4 | 97.4 | 98.2 | 99.4 | 98.2 | 99.9 | 111.4 |
| 4-0-6 | 86.4 | 88.0 | 87.6 | 85.1 | 87.8 | 80.9 |
| 5-0-6 | 89.8 | 85.9 | 85.5 | 91.5 | 84.5 | 80.9 |
| 3-0-5 | 86.4 | 88.0 | 87.6 | 85.1 | 87.8 | 86.7 |

**Supplementary Table 17** | Crystallographic Data for AgBiS$_2$ in $P3_221$ at room temperature (298 K) from PND.

| Lattice constant $a$ (Å) | 4.06954(18) | | | | |
|---|---|---|---|---|---|
| Lattice constant $c$ (Å) | 19.0809(10) | | | | |
| Volume (Å$^3$) | 273.67(3) | | | | |
| Crystal System | Trigonal | | | | |
| Space Group | $P3_221$ (no. 154) | | | | |
| Number of Formula Units | 3 | | | | |
| Diffractometer | D2B Beamline (ILL) | | | | |
| Structure solution and Refinement Program Package | Topas Academic V 7.25 | | | | |
| $R_{wp}$ | 4.076 | | | | |
| Goodness of Fit | 11.654 | | | | |
| Atomic Coordinates and Isotropic Displacement Parameters (Å$^2$) | | | | | |
| Atom | Site | x | y | z | $B_{iso}$ (Å$^2$) |
| Ag | *3b* | 0.6651(16) | 0 | 1/6 | 3.16(14) |

| Bi | *3a* | 0.7056(7) | 0 | 2/3 | 1.04(7) |
| S | *6c* | 0.4350(14) | 0.350(4) | 0.7524(4) | 2.01(13) |

## Supplementary Note 3 | Optical analysis

### *Absorption profile and energetic disorder*

In order to acquire the absorption spectra, we pressed the powders into pellets with smooth surfaces. However, due to strong light scattering, it was still challenging to determine the absorption spectra from reflectance measurements. Therefore, we adopted photothermal deflection spectroscopy (PDS) to characterize the absorbance profile here instead (Fig. 4a). In PDS measurements, the absorbance profile of a sample is determined through the deflection of a probe beam when the sample absorbs light, which will not be affected by light scattering. As a result, PDS cannot offer absolute absorbance values, but can provide a very precise relative absorbance profile even in the sub-gap region. With the help of PDS measurements, we can obtain Tauc plots for all of our three AgBiS$_2$ samples, as shown in Supplementary Fig. 22. We extracted a similar bandgap of 0.79 eV and 0.82 eV for the cubic and cation-ordered phase, respectively. Note that the extracted bandgap of our cubic AgBiS$_2$ pellet is close to the value reported for the bulk sample in previous work[11] (~0.8 eV). On the other hand, no experimental bandgap has yet been reported for cation-ordered AgBiS$_2$ so far, while the value acquired here was in line with the calculated bandgap of the $P3_221$ phase AgBiS$_2$ (0.86 eV), as shown in Table 4.

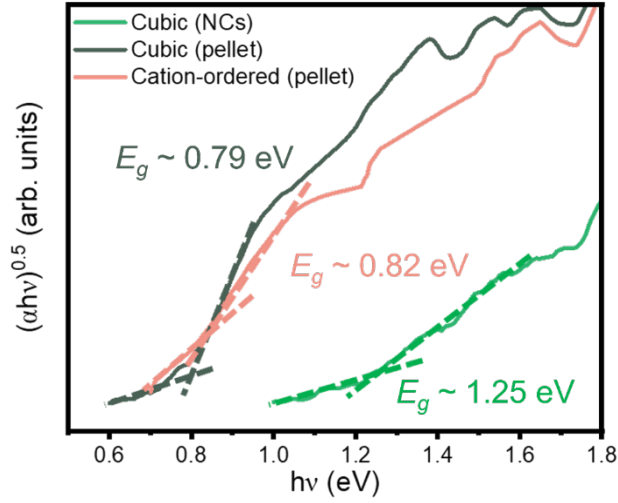

**Supplementary Fig. 22 | Tauc plots acquired from the measured PDS data** as displayed in Fig. 4a, along with the extracted bandgap values of three AgBiS$_2$ samples.

The sub-gap absorption tails were fitted using a bandgap fluctuation model[12], which was developed to account for energetic disorder (Fig. 4a). In this method, the electronic bandgap ($E_{\mathrm{g}}$) is expressed by a Gaussian probability distribution ($G(E_g)$) with an average bandgap $E_{\mathrm{g,m}}$ and width $\sigma_{\mathrm{g}}$. The absorbance ($A$) can then be written as

$$A \propto \int_0^\infty \frac{\sqrt{\hbar\omega - E_g}}{\hbar\omega}\, G(E_g)\ \mathrm{d}E_g \qquad (S3)$$

$$G(E_{\mathrm{g}}) = \frac{1}{\sigma_{\mathrm{g}}\sqrt{2\pi}} \exp\left(-\frac{\left(E_{\mathrm{g}} - E_{\mathrm{g,m}}\right)^2}{2\,\sigma_g^2}\right) \qquad (S4)$$

where $\hbar\omega$ is the photon energy. The fitted $\sigma_{\mathrm{g}}$ values are 0.165, 0.226 and 0.415 eV for cubic powders, cation-ordered powders and cubic NCs, respectively. Despite possessing similar degrees of cation disorder (refer to structural analysis section), there is a large discrepancy in $\sigma_g$ between the cubic powders (0.165 eV) and cubic NCs (0.415 eV), indicating that the cubic NCs are more energetically disordered than cubic powders. This is consistent with the much steeper absorption onset exhibited by the cubic powders than NCs. In contrast to a prior work that only analyzed NCs[13], our findings here suggest that the homogeneity of cation disorder may not be the decisive factor determining the electronic landscape of $AgBiS_2$. Cation-ordered powders exhibit a higher $\sigma_g$ value (0.226 eV) compared to the cubic powders, which could be caused by a higher defect density from a lower processing-temperature and reduced crystallinity.

### *Origins of the ultrafast decay in $AgBiS_2$ NCs*

An ultrafast ps decay in the OPTP signal could also arise due to exciton formation, which reduces the charge-to-photon branching ratio. However, many prior studies have found the cubic $AgBiS_2$ NCs to have free carriers rather than excitons[13, 14], and we synthesized our NCs in accordance with these prior reports. We do observe a higher-energy absorption onset in the NC samples (Supplementary Fig. 22), but this cannot be unambiguously assigned to quantum confinement, since prior works that varied the size of NCs did not observe a significant change in bandgap[15, 16]. Indeed, the powders may be observed to absorb further into the infrared simply by having more material present than the thin NCs. A decrease in the charge-to-photon branching ratio can also be obtained through non-radiative recombination via defects, but a decay on a ps timescale requires an unrealistically high defect density[17]. Therefore, we assign the ultrafast decay in the cubic NCs OPTP signal to carrier localization, and the slower decays in the powder samples to more band-like transport.

***OPTP Two-Level Model Analysis***

To quantify charge-carrier localization in cubic AgBiS$_2$ NCs, the OPTP transient was fitted by the two-level mobility model proposed by Buizza et al[18]. In this model, two different states contribute to the overall photoconductivity: a delocalized state (with a delocalized mobility $\mu_{deloc}$) which is initially populated after photoexcitation and a localized state (with a localized mobility $\mu_{loc}$) which is populated through the carrier localization process at a localization rate $k_{loc}$, followed by a slow relaxation step to the ground state with a recombination rate $k_1$. The overall $\frac{\Delta T}{T}$ signal then can be described as:

$$\frac{\Delta T}{T} = \frac{-n_0 e d}{\epsilon_0 c(n_A + n_B)}\left(\left(\mu_{deloc} - \frac{\mu_{loc}k_{loc}}{k_{loc} - k_1}\right)e^{-k_{loc}t} + \frac{\mu_{loc}k_{loc}}{k_{loc} - k_1}e^{-k_1 t}\right) \qquad (S5)$$

where $n_0$ and $d$ represent initial charge-carrier density after photoexcitation and film thickness, respectively. $n_A$ = 1 for vacuum and $n_B$ = 2.13 for a z-cut quartz substrate. To account for the Instrument Response Function, Equation S5 is further convolved with a normalised Gaussian function with broadening $\sigma_{IRF}$ = 100 $fs$ as described in Ref. 18.

***Modified Drude-Smith Model***

The modified Drude-Smith (mDS) Model is used here, since it can phenomenologically describe carrier transport within specific domains[19]. This model is expressed as:

$$\mu_{mDS}(f) = \mu_0(f)\left(1 - \frac{c}{1 - i2\pi f \tau''}\right) \qquad (S6)$$

with $\mu_0$ the classical Drude mobility, $f$ the probing frequency of the applied electric field, and $c$ a localization strength parameter, which varies from 0 (Drude-like carriers) and 1 (fully localized carriers). $\tau''$ can be further represented as:

$$\tau'' = \frac{eL_{loc}^2}{12k_BT\mu_0(f=0)} \qquad (S7)$$

with $e$, $k_B$, $T$ the elementary charge, Boltzmann constant, and temperature, respectively. $L_{loc}$ refers to the domain size that a carrier is localized.

***OPTP Probing Length***

The OPTP probing length ($L_P$) can be estimated as[19]:

$$L_P \approx \sqrt{\frac{6k_BT\mu}{\pi q f}} \qquad (S8)$$

where $k_B$, $T$, $f$ and $\mu$ denotes Boltzmann constant, temperature, probing frequency of the applied electric field, and the sum mobility, respectively. $L_P$ is calculated to be 0.4 nm for cubic NCs, much smaller than NC size. However, $L_P$ of both cubic and cation-ordered powders cannot be accurately acquired due to their unknown thicknesses and hence inaccurate mobilities.

## Supplementary Note 4 | Computational analysis of polarons in AgBiS₂

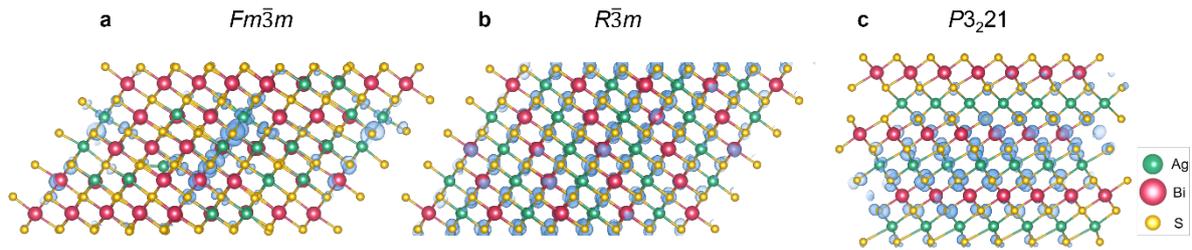

**a** $Fm\overline{3}m$  **b** $R\overline{3}m$  **c** $P3_221$

Ag
Bi
S

**Supplementary Fig. 23 | Calculated charge density isosurfaces of holes** (blue regions) for the **a** $Fm\overline{3}m$, **b** $R\overline{3}m$ and **c** $P3_221$ phases. The green, red, and yellow spheres refer to Ag, Bi, and S ions, respectively.

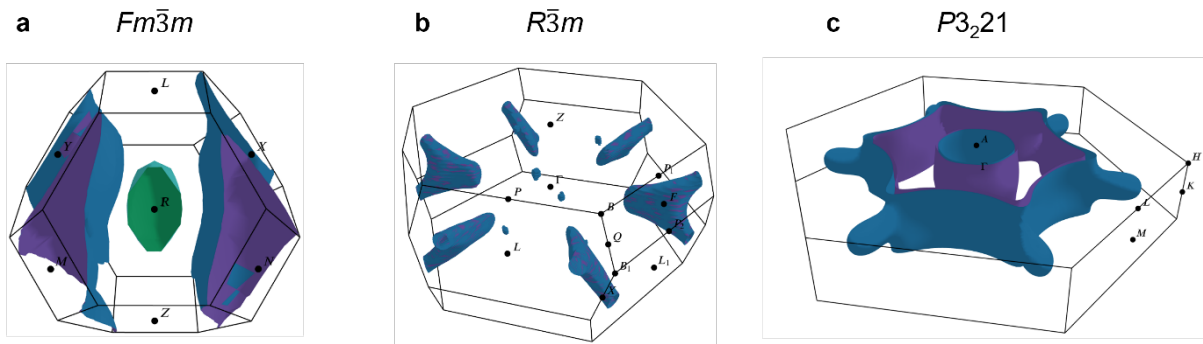

**a** $Fm\overline{3}m$  **b** $R\overline{3}m$  **c** $P3_221$

**Supplementary Fig. 24 | Calculated Fermi surface at the valence band** (0.1 eV below the VBM) for the **a** $Fm\overline{3}m$, **b** $R\overline{3}m$ and **c** $P3_221$ phases.

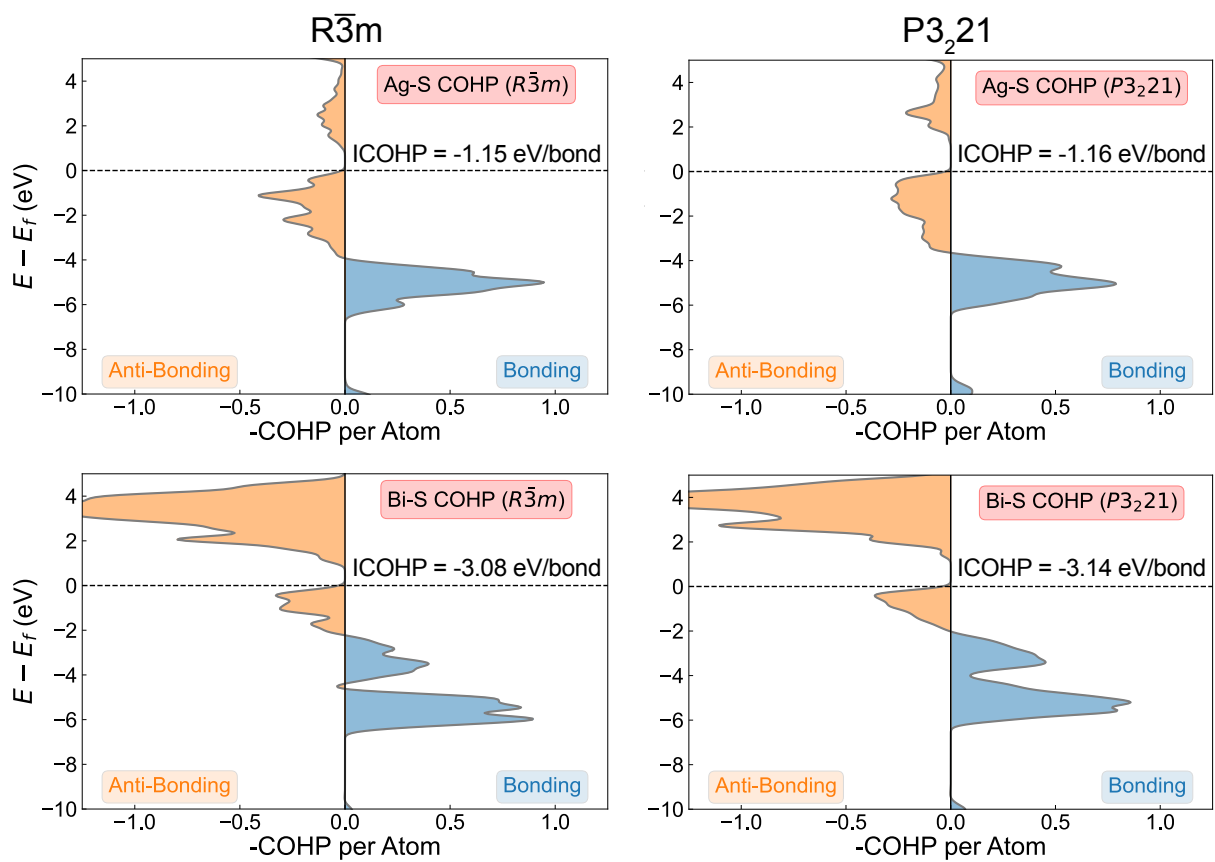

**Supplementary Fig. 25 | Crystal Orbital Hamilton Population (COHP) plots** for $R\bar{3}m$ (left) and $P3_221$ (right) phases of AgBiS$_2$, with the corresponding integrated COHP (ICOHP) values given alongside, showing that the $C2/c$ phase results in greater Ag-S and Bi-S bonding interactions. Plots generated using LOBSTER[20] and pymatgen[21]. 0.5 eV Gaussian broadening is applied.

## Supplementary Note 5 | Energetic disorder and OPTP of AgBiS₂ thin films

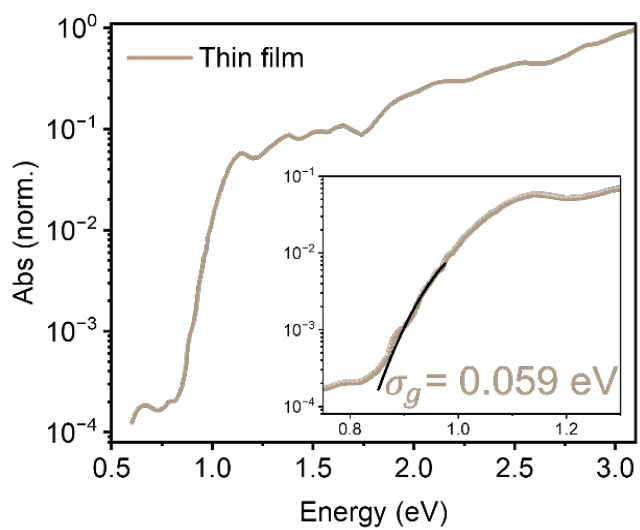

**Supplementary Fig. 26 | Absorbance spectra of thin film cubic-phase AgBiS₂** from photothermal deflection spectroscopy. Inset is the bandgap fluctuation model fitting showing the extract $\sigma_g$ values of thin film AgBiS₂ as 0.059 eV.

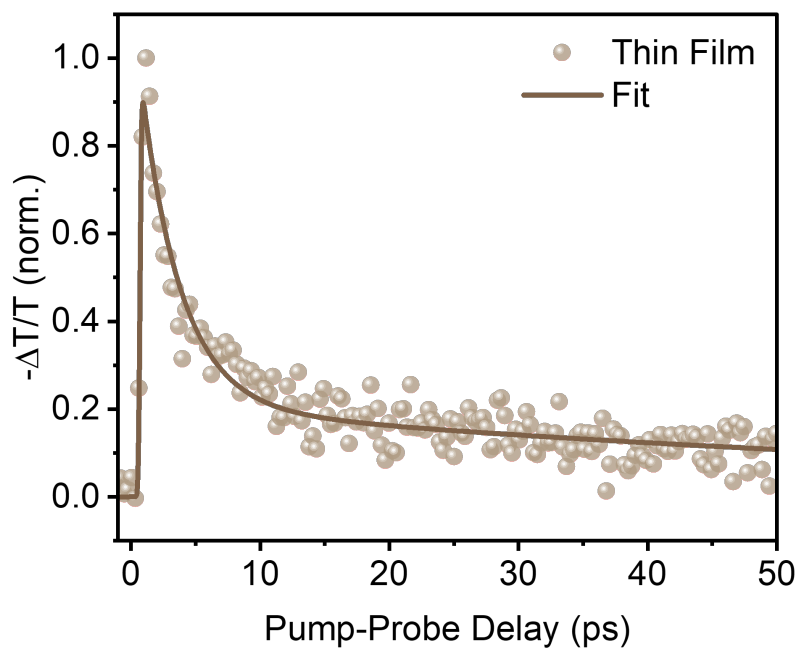

**Supplementary Fig. 27 | OPTP transients of thin film cubic AgBiS₂.**